\documentclass[thm-restate]{lmcs} 
\pdfoutput=1
\usepackage[utf8]{inputenc}

\usepackage{lastpage}
\lmcsdoi{19}{4}{19}
\lmcsheading{}{\pageref{LastPage}}{}{}%
{Aug.~16,~2022}{Dec.~05,~2023}{}


\usepackage{amssymb}
\usepackage{hyperref}
\usepackage{xspace}
\usepackage{colortbl}
\usepackage{cleveref}

\usepackage{tikz}
\usetikzlibrary{automata,positioning,fit,calc,shapes.geometric}

\newcommand{\tI}{2^{I}}
\newcommand{\tO}{2^{O}}
\newcommand{\sI}{\Sigma_{I}}
\newcommand{\sO}{\Sigma_{O}}

\newcommand{\tup}[1]{\left< #1\right>}
\newcommand{\ST}{\, :\, }

\newcommand{\cA}{\mathcal{A}}
\newcommand{\cB}{\mathcal{B}}

\newcommand{\cN}{\mathcal{N}}
\newcommand{\cD}{\mathcal{D}}
\newcommand{\cT}{\mathcal{T}}
\newcommand{\cP}{\mathcal{P}}
\newcommand{\cS}{\mathcal{S}}

\newcommand{\bbB}{\mathbb{B}}
\newcommand{\bbN}{\mathbb{N}}

\newcommand{\DFA}{\mbox{\rm DFA}\xspace}
\newcommand{\NFA}{\mbox{\rm NFA}\xspace}

\newcommand{\NFAs}{\mbox{\rm NFAs}\xspace}

\newcommand{\PA}{\mbox{\rm PA}\xspace}

\newcommand{\lab}{\boldsymbol{\ell}}

\newcommand{\fL}{\Lambda}
\newcommand{\fP}{\mathfrak{P}}

\renewcommand{\vec}[1]{\boldsymbol{#1}}
\newcommand{\req}[1][]{\asymp_{#1}}
\newcommand{\lcm}{{\rm lcm}}
\newcommand{\runs}[2]{\stackrel{#1}{\longrightarrow}_{#2}}
\newcommand{\tr}{{\mathtt{Tr}}}
\newcommand{\perm}{{\mathtt{Perm}}}

\newcommand{\WRT}{w.r.t.\! }
\newcommand{\condset}[2]{\left\{\,#1\ \middle|\ #2\,\right\}}

\newtheorem{text-algorithm}[thm]{Algorithm}
\newtheorem*{text-algorithm*}{Algorithm}

\crefname{obs}{Observation}{Observations} 
\crefname{thm}{Theorem}{Theorems} 
\crefname{lem}{Lemma}{Lemmas} 
\crefname{rem}{Remark}{Remarks} 
\crefname{exa}{Example}{Examples} 
\crefname{cor}{Corollary}{Corollaries} 

\newcommand{\coNPSPACE}{\textsf{coNPSPACE}}
\newcommand{\PSPACE}{\textsf{PSPACE}}
\newcommand{\EEEEEXP}{\textsf{5-EXP}}

\begin{document}
\bibliographystyle{alphaurl}

\title{Simulation by Rounds of Letter-to-Letter Transducers}

\titlecomment{{\lsuper*}A preliminary version was published in CSL 2022~\cite{AA22}}
\thanks{This research was supported by the ISRAEL SCIENCE FOUNDATION (grant No. 989/22)}	

\author[A.~Abu Nassar]{Antonio Abu Nassar}[]
\author[S.~Almagor]{Shaull Almagor\lmcsorcid{0000-0001-9021-1175}}[]

\address{Technion, Israel}	
\email{antonio@cs.technion.ac.il, shaull@technion.ac.il}  




\begin{abstract}
Letter-to-letter transducers are a standard formalism for modeling reactive systems. Often, two transducers that model similar systems differ locally from one another, by behaving similarly, up to permutations of the input and output letters within ``rounds''.
In this work, we introduce and study notions of simulation by rounds and equivalence by rounds of transducers. In our setting, words are partitioned to consecutive subwords of a fixed length $k$, called rounds. Then, a transducer $\cT_1$ is $k$-round simulated by transducer $\cT_2$ if, intuitively, for every input word $x$, we can permute the letters within each round in $x$, such that the output of $\cT_2$ on the permuted word is itself a permutation of the output of $\cT_1$ on $x$.
Finally, two transducers are $k$-round equivalent if they simulate each other.

We solve two main decision problems, namely whether $\cT_2$ $k$-round simulates $\cT_1$ (1) when $k$ is given as input, and (2) for an existentially quantified $k$.

We demonstrate the usefulness of the definitions by applying them to process symmetry: a setting in which a permutation in the identifiers of processes in a multi-process system naturally gives rise to two transducers, whose $k$-round equivalence corresponds to stability against such permutations.
\end{abstract}

\maketitle

\section{Introduction}
\label{sec:introduction}
Reactive systems interact with their environment by receiving inputs, corresponding to the state of the environment, and sending outputs, which describe actions of the system. Finite-state reactive systems are often modeled by \emph{transducers} -- finite-state machines over alphabets $\sI$ and $\sO$ of inputs and outputs, respectively, which read an input letter in $\sI$, and respond with an output in $\sO$.
Such transducers are amenable to automatic verification of certain properties (e.g., LTL model-checking), and are therefore useful in practice. Nonetheless, modeling complex systems may result in huge transducers, which makes verification procedures prohibitively expensive, and makes understanding the constructed transducers difficult.

A common approach to gain a better understanding of a transducer (or more generally, any system) is \emph{simulation}~\cite{Milner1971}, whereby a transducer $\cT_1$ is simulated by a ``simpler'' transducer $\cT_2$ in such a way that model checking is easier on $\cT_2$, and the correctness of the desired property is preserved under the simulation. Usually, ``simpler'' means smaller, as in standard simulation~\cite{Milner1971} and fair simulation~\cite{Henzinger1997}, but one can also view e.g., linearization of concurrent programs~\cite{Herlihy1987} as a form of simulation by a simpler machine.

In this work, we introduce and study new notions of simulation and of equivalence for transducers, based on \emph{rounds}: consider an input word $x\in \sI^*$ whose length is $k\cdot R$ for some $k,R>0$. We divide the word into $R$ disjoint infixes of length $k$, each called a round of $w$. We then say that two words $x,x'\in \sI^{kR}$ are $k$-round equivalent, denoted $x'\req[k]x$, if $x'$ is obtained from $x$ by permuting the positions of letters within each round of $x$. For example $abcabc$ and $cbaacb$ are $3$-round equivalent, since $cba$ is a permutation of $abc$ and so is $acb$. \Cref{example:word-req} presents a pair of words that are 3-round equivalent but not 4-round equivalent. We now say that a transducer $\cT_1$ is \emph{$k$-round simulated} by a transducer $\cT_2$, denoted $\cT_1\prec_{k} \cT_2$, if for every\footnote{Our formal definition allows to also restrict the input to some regular language $\fL\subseteq \sI^*$, see~\cref{sec:round-equivalence}.} input $x\in \sI^{kR}$ we can find $x'\req[k]x$ such that the outputs of $\cT_1$ on $x$ and $\cT_2$ on $x'$, denoted $y,y'$ respectively, are also round equivalent: $y'\req[k] y$.
Intuitively, $\cT_1\prec_{k}\cT_2$ means that every behaviour of $\cT_1$ is captured by $\cT_2$, up to permutations within each round.
When we have both $\cT_1\prec_{k}\cT_2$ and $\cT_2\prec_{k}\cT_1$, we say that they are $k$-round equivalent and denote this by $\cT_1\equiv_{k}\cT_2$.

The benefit of $k$-round simulation is twofold. First, it may serve as an alternative simulation technique for reducing the state space while maintaining the correctness of certain properties. Second, we argue that $k$-round simulation is in and of itself a design concern. Indeed, in certain scenarios, as follows, we can naturally design a transducer $\cT_2$ that performs a certain task in an ideal, but not realistic, way, and we want to check that an existing design, namely $\cT_1$, is simulated by this ideal. In particular, this is useful when dealing with systems that naturally work in rounds, such as schedulers (e.g., Round Robin, cf.~\cref{example:transducer-req}), arbiters, and other resource allocation systems.

We now demonstrate both benefits by an example.
\begin{exa}
	\label{example:MC_rounds}
	Consider a monitor $M$ for the fairness of a distributed system with $10$ processes $\cP=\{1,\ldots,10\}$. At each timestep, $M$ receives as input the ID of the process currently working. The monitor then verifies that in each round of $10$ steps, every process works exactly once. As long as this holds, the monitor keeps outputting $\texttt{safe}$; otherwise, it outputs $\texttt{error}$.

	$M$ can be modeled by a transducer $\cT_1$ that keeps track of the set of processes that have worked in the current round. Thus, the transducer has at least $2^{10}$ states, as it needs to keep track of the subset of processes that have been seen.

	It is not hard to see that $\cT_1$ is $10$-round simulated by an ``ideal'' transducer $\cT_2$ which expects to see the processes in the order $1,\ldots,10$. This transducer needs roughly $10$ states, as it only needs to know the index of the next process it expects to see.

	Now, suppose we want to verify some correctness property which is invariant to permutations of the processes within each round of length 10, such as ``if there is no \texttt{error}, then Process $3$ works at least once every 20 steps''. Then we can verify this against the much smaller $\cT_2$.
\end{exa}

The notion of $k$-round simulation arises naturally in the setting of \emph{process symmetry}. There, the input and output alphabets are $\sI=\tI$ and $\sO=\tO$ respectively, where $I=\{i_1,\ldots,i_m\}$ and $O=\{o_1,\ldots,o_m\}$ represent signals corresponding to $m$ processes. Process symmetry addresses the scenario where the identifiers of the processes may be scrambled. For example, if the input $\{i_1,i_2\}$ is generated, the system might actually receive an input $\{i_7,i_4\}$. A system exhibits process symmetry if, intuitively, its outputs are permuted in a similar way to the inputs. Unfortunately, deterministic systems that are process symmetric are extremely naive, as process symmetry is too restrictive for them. While this can be overcome using probabilistic systems, as studied in~\cite{Almagor2020b}, it is also desirable to find a definition that is suited for deterministic systems. As we show in~\cref{sec:application}, $k$-round simulation provides such a definition.

The main contributions of this work are as follows. We introduce the notion of $k$-round simulation and $k$-round equivalence, and define two decision problems pertaining to them: in \emph{fixed round simulation} we need to decide whether $\cT_1\prec_k \cT_2$ for a given value of $k$, and in \emph{existential round simulation} we need to decide whether there exists some value of $k$ for which $\cT_1\prec_k \cT_2$ holds. In fact, we consider a somewhat more elaborate setting, by also allowing the inputs to $\cT_1$ to be restricted to some regular language $\fL$.
We solve the first problem by reducing it to the containment of two nondeterministic automata. For the second problem, things become considerably more difficult, and the solution requires several constructions, as well as tools such as Presburger arithmetic and Parikh's theorem.
In addition, we demonstrate the usefulness of the definitions in relation to process symmetry.

\subsection*{Related Work}
Simulation relations between systems are a well studied notion. We refer the reader to~\cite[Chapter 13]{Clarke2018a} and references therein for an exposition. The connection of our notion with standard simulation is only up to motivation, as our measure is semantic: it does not directly relate to the state space; instead, it refers to the \emph{behaviour} of the system rather than its structure.

On the technical level, our work is closely related to \emph{commutative automata}~\cite{Brzozowski1973} and \emph{jumping automata}~\cite{Fernau2015,Meduna2012} --- models of automata capable of reading their input in a discontinuous manner, by jumping from one letter to another. Indeed, our notion of round simulation essentially allows the simulating transducer to read the letters within rounds in a discontinuous manner. This similarity is manifested implicitly in~\cref{sec:proof_of_bound}, where we encounter similar structures as e.g. the commutative closure in~\cite{Hoffmann2020} (although the analysis here has a different purpose).

Finally, the initial motivation for this work comes from \emph{process symmetry}~\cite{Almagor2020b,Clarke1996,Emerson1996,Ip1996,Lin2016}. We explore the connections in depth in~\cref{sec:application}.

\subsection*{Paper Organization}
The rest of this work is organized as follows.
In~\cref{sec:preliminaries} we present some basic definitions used throughout the paper. In~\cref{sec:round-equivalence} we introduce $k$-round simulation and equivalence, define the relevant decision problems, and study some fundamental properties of the definitions. In~\cref{sec:deciding_fixed_round_sim} we solve fixed round simulation, while developing some technical tools and characterizations that are reused later. \cref{sec:deciding_existential_round_sim} is our main technical result, where we develop a solution for existential round simulation. In particular, in~\cref{sec:intuitive_overview} we give an overview of the solution, before going through the technical details in~\cref{sec:proof_of_bound}. In~\cref{sec:lower_bounds_existential} we give lower bounds for the existential setting.
In~\cref{sec:application} we use round simulation to obtain a definition of process symmetry for deterministic transducers, along with an algorithm for deciding it. In~\cref{sec:equiv_mapping} we study the mapping between transducers that induces a simulation.
In~\cref{sec:other_notions} we study variants of symmetry and simulation, both refining and coarsening the previous notions. Finally, we conclude with some open problems in~\cref{sec:discussion}.

\section{Preliminaries}
\label{sec:preliminaries}

\paragraph*{Automata}
A \emph{deterministic finite automaton (\DFA)} is $\cA=\tup{\Sigma,Q,q_0,\delta,F}$, where $Q$ is a finite set of states, $q_0 \in Q$ is an initial state, $\delta: Q\times \Sigma \to Q$ is a transition function, and $F\subseteq Q$ is the set of accepting states.

The \emph{run} of $\cA$ on a word $w=\sigma_0 \cdot \sigma_1 \cdots \sigma_{n-1}\in \Sigma^*$ is a sequence of states $q_0,q_1,\ldots,q_n$ such that $q_{i+1} = \delta(q_i,\sigma_{i})$ for all $0\le i<n$.
The run is \emph{accepting} if $q_n\in F$. A word $w \in \Sigma^*$ is \emph{accepted} by $\cA$ if the run of $\cA$ on $w$ is accepting. The \emph{language} of $\cA$, denoted $L(\cA)$, is the set of words that $\cA$ accepts.
We also consider \emph{nondeterministic finite automata (\NFA)}, where $\delta:Q\times \Sigma\to 2^Q$ and there can be multiple initial states. Then, a run of $\cA$ on a word $w\in \Sigma^*$ as above is a sequence of states $q_0,q_1,\ldots,q_n$ such that $q_0$ is an initial state and $q_{i+1} \in \delta(q_i,\sigma_{i})$ for all $0\le i<n$. Analogously to the deterministic setting, the language of $\cA$ is the set of words that have an accepting run.
We denote by $|\cA|$ the number of states of $\cA$.

As usual, we denote by $\delta^*$ the transition function lifted to words.
For states $q,q'$ and $w\in \Sigma^*$, we write $q\runs{w}{\cA}q'$ if $q'\in \delta^*(q,w)$. That is, if there is a run of $\cA$ from $q$ to $q'$ while reading $w$.

\begin{figure}[ht]
	\centering
	\begin{tikzpicture}[shorten >=1pt,node distance=1cm and 2cm,on grid,auto]
		\node[state] (q_0) [initial above] {$q_0$};
		\node[state] (q_1) [right=of q_0] {$q_1$};
		\node[state] (q_2) [right=of q_1, accepting] {$q_2$};
		\path[->]
		(q_0) edge node {$a$} (q_1)
		(q_1) edge node {$b$} (q_2)
		(q_2) edge [loop above] node {$a,b$} (q_2);
	\end{tikzpicture}
	\caption{A nondeterministic automaton with one initial state $q_0$ and one accepting state $q_2$.}
	\label{fig:type}
\end{figure}
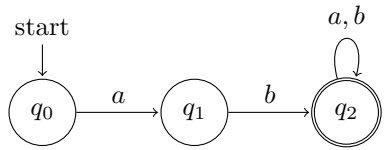

An \NFA $\cA$ can be viewed as a morphism from $\Sigma^*$ to the monoid $\bbB^{Q\times Q}$ of $Q\times Q$ Boolean matrices, where we associate with a letter $\sigma\in \Sigma$ its \emph{type} $\tau_{\cA}(\sigma)\in \bbB^{Q\times Q}$ defined by $(\tau_{\cA}(\sigma))_{q,q'}=1$ if $q\runs{\sigma}{\cA}q'$, and $(\tau_{\cA}(\sigma))_{q,q'}=0$ otherwise. We lift the definition of types to $\Sigma^*$ by defining, for a word $w=\sigma_1\cdots \sigma_n\in \Sigma^*$, its type as $\tau_{\cA}(w)=\tau_{\cA}(\sigma_1)\cdots \tau_{\cA}(\sigma_n)$ where the concatenation denotes Boolean matrix product. It is easy to see that $(\tau_{\cA}(w))_{q,q'}=1$ iff $q\runs{w}{\cA}q'$.
For example, the types of the letters $a$ and $b$ in the automaton in~\cref{fig:type} are the $3\times 3$ matrices
\[
	\tau_\cA(a)=
	\begin{array}{r@{}c}
		& \begin{array}{c c c} q_0 & q_1 & q_2 \end{array} \\
		\begin{array}{r}
			q_0 \\
			q_1 \\
			q_2 \\
		\end{array}
		& \left[ \begin{array}{ c c c }
					 0 & 1 & 0 \\
					 0 & 0 & 0 \\
					 0 & 0 & 1 \\
		\end{array} \right]
	\end{array}, \quad
	\tau_\cA(b)=
	\begin{array}{r@{}c}
		& \begin{array}{c c c} q_0 & q_1 & q_2 \end{array} \\
		\begin{array}{r}
			q_0 \\
			q_1 \\
			q_2 \\
		\end{array}
		& \left[ \begin{array}{ c c c }
					 0 & 0 & 0 \\
					 0 & 0 & 1 \\
					 0 & 0 & 1 \\
		\end{array} \right]
	\end{array},
\]
and the type of the word $w=ab$ in the transducer in~\cref{fig:type} is the matrix
\[
	\tau_\cA(w)=
	\begin{array}{r@{}c}
		& \begin{array}{c c c} q_0 & q_1 & q_2 \end{array} \\
		\begin{array}{r}
			q_0 \\
			q_1 \\
			q_2 \\
		\end{array}
		& \left[ \begin{array}{ c c c }
					 0 & 0 & 1 \\
					 0 & 0 & 0 \\
					 0 & 0 & 1 \\
		\end{array} \right]
	\end{array}
	=\tau_\cA(a)\cdot \tau_\cA(b).
\]

\paragraph*{Transducers}
Consider two sets $\sI$ and $\sO$ representing input and output alphabets, respectively. A \emph{$\sI/\sO$ transducer} is $\cT=\tup{\sI,\sO,Q,q_0,\delta,\lab}$ where $Q$, $q_0\in Q$, and $\delta:Q\times \sI\to Q$ are as in a \DFA, and $\lab:Q\to \sO$ is a labelling function on the states. For a word $w\in \sI^*$, consider the run $\rho=q_0,\ldots,q_n$ of $\cT$ on $w$. We define its output $\lab(\rho)=\lab(q_1)\cdots\lab(q_n)\in \sO^*$, and we define the output of $\cT$ on $w$ to be $\cT(w)=\lab(\rho)$. Observe that we ignore the labelling of the initial state in the run, so that the length of the output matches that of the input.

\paragraph*{Words and rounds}
Consider a word $w=\sigma_0\cdots \sigma_{n-1}\in \Sigma^*$. We denote its length by $|w|$, and for $0\le i\le j< |w|$ we define $w[i:j]=\sigma_i\cdots \sigma_j$.
For $k>0$, if $|w|=kR$ for some $R\in \bbN$, then for every $0\le r<R$ we refer to $w[rk:r(k+1)-1]$ as the \emph{$r$-th round} in $w$ (of length $k$), and we write $w=\gamma_0\cdots \gamma_{R-1}$ where $\gamma_r$ is the $r$-th round. We emphasize that $k$ indicates the length of each round, not the number of rounds.

In particular, throughout the paper we consider words $(x,y)\in (\sI^k\times \sO^k)^*$ and their rounds of length $k$. In such cases, we sometimes use the natural embedding of $(\sI^k\times \sO^k)^*$ in $(\sI\times \sO)^*$ and in $\sI^*\times \sO^*$, and refer to these sets interchangeably.

\paragraph*{Parikh vectors and permutations}
Consider an alphabet $\Sigma$. For a word $w\in \Sigma^*$ and a letter $\sigma\in \Sigma$, we denote by $\#_\sigma(w)$ the number of occurrences of $\sigma$ in $w$.
The \emph{Parikh map} $\fP:\Sigma^*\to \bbN^\Sigma$ maps every word $w\in \Sigma^*$ to a \emph{Parikh vector} $\fP(w)\in \bbN^\Sigma$, where $\fP(w)(\sigma)=\#_\sigma(w)$. We lift this to languages by defining, for $L\subseteq \Sigma^*$, $\fP(L)=\{\fP(w):w\in L\}$.

For $\vec{p}\in \bbN^\Sigma$ (in the following we consistently denote vectors in $\bbN^\Sigma$ by bold letters) we write $|\vec{p}|=\sum_{\sigma\in \Sigma}\vec{p}(\sigma)$. In particular, for a word $w\in \Sigma^*$ we have $|\fP(w)|=|w|$.

By Parikh's theorem~\cite{Parikh1966}, for every \NFA $\cA$ we have that $\fP(L(\cA))$ is a \emph{semilinear set} -- that is, a finite union of sets of the form $\condset{\vec{p}+\lambda_1 \vec{s_1}+\ldots+\lambda_1 \vec{s_m}}{\lambda_1,\ldots,\lambda_m\in \bbN}$ where $\vec{p},\vec{s_1},\ldots,\vec{s_m}\in \bbN^d$.

Consider words $x,y\in \Sigma^*$. We say that $x$ is a \emph{permutation} of $y$ if $\fP(x)=\fP(y)$ (indeed, in this case $y$ can be obtained from $x$ by permuting its letters). In particular this implies $|x|=|y|$.

\section{Round Simulation and Round Equivalence}
\label{sec:round-equivalence}
Consider two $k$-round words $x,y\in \Sigma^{kR}$ with the same number of rounds $R$, and denote their rounds by $x=\alpha_0\cdots \alpha_{R-1}$ and $y=\beta_0\cdots \beta_{R-1}$. We say that $x$ and $y$ are \emph{$k$-round equivalent}, denoted $x\req[k] y$ (or $x\req y$, when $k$ is clear from context)\footnote{Conveniently, our symbol for round equivalence is a rounded equivalence.}, if for every $0\le r<R$ we have that $\fP(\alpha_r)=\fP(\beta_r)$. That is, $x\req y$ iff the $r$-th round of $y$ is a permutation of the $r$-th round of $x$, for every $r$. Indeed, $\req$ is an equivalence relation.

\begin{exa}[Round-equivalence for words]
	\label{example:word-req}
	Consider the words $x=abaabbabbbaa$ and $y=baabbaabbaba$ over the alphabet $\Sigma=\{a,b\}$. Looking at the words as $3$-round words, one can see in~\cref{tab:word-req-3} that rounds of length 3 in $y$ are all permutations of those in $x$, which gives $x\req[3] y$. However, looking at the rounds of length 4 of $x,y$, the number of occurrences of $b$ already in the first round of $x$ and of $y$ is different, so $x\not\req[4] y$, as illustrated in~\cref{tab:word-req-4}.
	\begin{table}[!htb]
		\hspace{.05\linewidth}%
		\begin{minipage}{.4\linewidth}
			\centering
			\caption{$x$ and $y$ are 3-round equivalent}
			\vspace{2mm}
			\begin{tabular}{c||c|c|c|c}
				$x$ & aba & abb & abb & baa \\
				\hline
				$y$ & baa & bba & abb & aba \\
			\end{tabular}
			\label{tab:word-req-3}
		\end{minipage}%
		\hspace{.1\linewidth}%
		\begin{minipage}{.4\linewidth}
			\centering
			\caption{$x$ and $y$ are not 4-round equivalent}
			\vspace{2mm}
			\begin{tabular}{c||c|c|c}
				$x$ & \emph{abaa} & bbab & bbaa \\
				\hline
				$y$ & \emph{baab} & baab & baba \\
			\end{tabular}
			\label{tab:word-req-4}
		\end{minipage}
	\end{table}
\end{exa}

Let $\sI$ and $\sO$ be input and output alphabets, let $\fL\subseteq \sI^*$ be a regular language, and let $k>0$. Consider two $\sI/\sO$ transducers $\cT_1$ and $\cT_2$. We say that \emph{$\cT_2$ $k$-round simulates $\cT_1$ restricted to $\fL$}, denoted $\cT_1\prec_{k,\fL} \cT_2$, if for every $k$-round word $x\in \fL$ there exists a $k$-round word $x'\in \sI^*$ such that $x\req[k] x'$ and $\cT_1(x)\req[k] \cT_2(x')$.

Intuitively, $\cT_1\prec_{k,\fL} \cT_2$ if for every input word $x\in \fL$, we can permute each round of length $k$ in $x$ to obtain a new word $x'$, such that the outputs of $\cT_1$ on $x$ and of $\cT_2$ on $x'$ are $k$-round equivalent.
Note that the definition is not symmetric: the input $x$ for $\cT_1$ is universally quantified, while $x'$ is chosen according to $x$. We illustrate this in~\cref{example:def-asymmetric}.

If $\cT_1\prec_{k,\fL} \cT_2$ and $\cT_2\prec_{k,\fL} \cT_1$ we say that $\cT_1$ and $\cT_2$ are \emph{$k$-round equivalent restricted to $\fL$}, denoted $\cT_1\equiv_{k,\fL} \cT_2$.
In the special case where $\fL=\sI^*$ (i.e., when we require the simulation to hold for every input), we omit it from the subscript and write $\cT_1\prec_k \cT_2$.

\begin{rem}[On the role of $\fL$]
    \label{rmk:restriction_language}
    Transducers have a ``universal'' flavor, in that every input string is assigned an output. In many settings, however, inputs of interest should comply to some simple form, and are otherwise irrelevant. The restriction language $\fL$ allows the designer to specify that we only care about symmetry when the input is correctly formed.

    We note that it is technically easy to add a similar restriction-language for $\cT_2$, although we find it less motivated, as $\cT_2$ is meant to be an abstraction of $\cT_1$ for the purpose of verification, rather than a concrete model to act in an environment.
\end{rem}

\begin{exa}[Round Robin]
	\label{example:transducer-req}
	We consider a simple version of the Round Robin scheduler for three processes $\cP=\{0,1,2\}$. In each time step, the scheduler outputs either a singleton set containing the ID of the process whose request is granted, or an empty set if the process whose turn it is did not make a request.
	Depending on the ID $i\in \{0,1,2\}$ of the first process, we model the scheduler as a $2^{\cP}/2^{\cP}$ transducer $\cT_i = \tup{2^\cP, 2^\cP, Q, q_{(i-1)\%3}, \delta, \lab}$ depicted in~\cref{fig:RR}, where \% is the $\bmod$ operator, $Q=\left\{q_0,q_1,q_2,q'_0,q'_1,q'_2\right\}$, $\delta(q_i, \sigma)=q_{(i+1)\%3}$ if $i+1\in\sigma$ and $\delta(q_i,\sigma)=q'_{(i+1)\%3}$ otherwise, $\lab(q_i)=\{i\}$ and $\lab(q'_i)=\emptyset$.
	\begin{figure}[ht]
		\centering
		\small
		\begin{tikzpicture}[shorten >=1pt,node distance=1.5cm and 4cm,on grid,auto]
			\tikzset{state/.style={draw,ellipse,minimum size=0pt}};
			\clip (-1.5,1.5) rectangle (9, -3);
			\node[state] (q_0) {$q_0/{\color{red}\{0\}}$};
			\node[state] (q_1) [right=of q_0] {$q_1/{\color{red}\{1\}}$};
			\node[state] (q_2) [right=of q_1] {$q_2/{\color{red}\{2\}}$};
			\node[state] (q'_0) [below=of q_0] {$q'_0/{\color{red}\emptyset}$};
			\node[state] (q'_1) [below=of q_1] {$q'_1/{\color{red}\emptyset}$};
			\node[state] (q'_2) [below=of q_2] {$q'_2/{\color{red}\emptyset}$};
			\path[->]
			(q_0)  edge node {$1$} (q_1)
			edge [sloped] node[pos=0.7] {$\lnot 1$} (q'_1)
			(q_1)  edge node {$2$} (q_2)
			edge [sloped] node [pos=0.7] {$\lnot 2$} (q'_2)
			(q_2)  edge [bend right=15, above] node {$0$} (q_0)
			edge [sloped, out=160, in=130, looseness=1.5, above] node  [below,pos=0.5] {$\lnot 0$} (q'_0)
			(q'_0) edge [sloped] node [pos=0.7] {$1$} (q_1)
			edge node {$\lnot 1$} (q'_1)
			(q'_1) edge [sloped] node [pos=0.7] {$2$} (q_2)
			edge node {$\lnot 2$} (q'_2)
			(q'_2) edge [sloped, out=200, in=230, looseness=1.5, below] node [above,pos=0.5] {$0$} (q_0)
			edge [bend left=15] node {$\lnot 0$} (q'_0)
			(q_2);
		\end{tikzpicture}
		\caption{The transducer $\cT_i$ for RR, initial state omitted. The input letters $\sigma$ and $\lnot\sigma$ mean all letters from $2^\cP$ that, respectively, contain or do not contain $\sigma$. The labels are written in red.}
		\label{fig:RR}
	\end{figure}
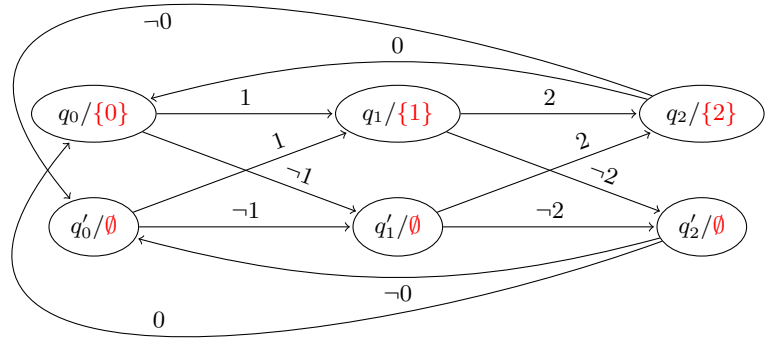

	Technically, the initial state changes the behaviour of $\cT_i$ significantly (e.g. we have $\cT_0(\{0\}\{2\}\{1\})=\{0\}\emptyset\emptyset$ whereas $\cT_1(\{0\}\{2\}\{1\})=\emptyset\{2\}\emptyset$). Conceptually, however, changing the initial state does not alter the behaviour, as long as the requests are permuted accordingly. This is captured by round equivalence, as follows.

	We argue that, if we allow permutation of the input letters, then the set of processes whose requests are granted in each round is independent of the start state. This is equivalent to saying $\cT_0 \equiv_3 \cT_j$ for $j\in \{1,2\}$, which indeed holds: if $j=1$ then we permute all rounds of the form $\sigma_0 \sigma_1 \sigma_2$ to $\sigma_1 \sigma_2 \sigma_0$, and similarly if $j=2$ then we permute all rounds to $\sigma_2 \sigma_0 \sigma_1$. It is easy to see that the run of $\cT_i$ on the permuted input grants outputs that are $3$-round equivalent to the output of $\cT_0$ on the non-permuted input.
\end{exa}

\begin{rem}
    \label{rmk:more_rounds_than_processes}
    In~\cref{example:transducer-req}, the constant $k$ of round equivalence is equal to the number of processes $k=3$. This need not be the case in general. Indeed, one could define Round Robin over 3 processes that follows the request order e.g., $111232332$. 
    It is easy to show that in this case, the natural round length is $9$, and that permutations of $3$-rounds are not enough to reorder inputs starting from different initial states.
\end{rem}

In~\cref{example:transducer-req}, the transducers satisfied not only round simulation, but also round equivalence. We now show that this is not always the case for simulating transducers.

\begin{exa}[Round simulation is not symmetric]
	\label{example:def-asymmetric}
	Consider the $\sI/\sO$ transducers $\cT_1$ and $\cT_2$ over the alphabet $\sI=\{a,b\}$ and $\sO=\{0,1\}$, depicted in~\cref{fig:def-asymmetry}.
	\begin{figure}[ht]
		\centering
		\begin{tikzpicture}[shorten >=1pt,node distance=1.1cm and 2cm,on grid,auto,initial text = {}]
			\clip (-0.9,0.55) rectangle (4.5, -1.5);
			\tikzset{every state/.style={minimum size=0pt}}
			\node[state,initial] (q_0) {\color{red} $1$};
			\node[state] (q_b) [right=of q_0] {\color{red} $0$};
			\node[state] (q_a) [below=of q_0] {\color{red} $0$};
			\node[state] (q_sink) [below=of q_b] {\color{red} $1$};
			\node[state] (q_bb) [right=of q_b] {\color{red} $1$};
			\path[->]
			(q_0)  edge [bend left=15] node {$b$} (q_b)
			edge [bend left=15] node {$a$} (q_a)
			(q_b)  edge [bend left=15] node {$a$} (q_0)
			edge node {$b$} (q_bb)
			(q_bb) edge node[pos=0.3] {$a,b$} (q_sink)
			(q_a)  edge node[below] {$a$} (q_sink)
			edge [bend left=15] node {$b$} (q_0)
			(q_sink) edge [loop right] node {$a,b$} ();
		\end{tikzpicture}%
		\hspace{2cm}%
		\begin{tikzpicture}[shorten >=1pt,node distance=1.1cm and 2cm,on grid,auto,initial text = {}]
			\clip (-0.9,0.55) rectangle (4.5, -1.5);
			\tikzset{every state/.style={minimum size=0pt}}
			\node[state,initial] (q_0) {\color{red} $0$};
			\node[state] (q_b) [right=of q_0] {\color{red} $1$};
			\node[state] (q_a) [below=of q_0] {\color{red} $0$};
			\node[state] (q_sink) [below=of q_b] {\color{red} $1$};
			\node[state] (q_bb) [right=of q_b] {\color{red} $0$};
			\path[->]
			(q_0)  edge [bend left=15] node {$b$} (q_b)
			edge [bend left=15] node {$a$} (q_a)
			(q_b)  edge [bend left=15] node {$a$} (q_0)
			edge node {$b$} (q_bb)
			(q_bb) edge node[pos=0.3] {$a,b$} (q_sink)
			(q_a)  edge node[below] {$a$} (q_sink)
			edge [bend left=15] node {$b$} (q_0)
			(q_sink) edge [loop right] node {$a,b$} ();
		\end{tikzpicture}%
		\caption{Transducers $\cT_1$ (left) and $\cT_2$ (right) illustrate the asymmetry in the definition of round equivalence (see~\cref{example:def-asymmetric}).}
		\label{fig:def-asymmetry}
	\end{figure}
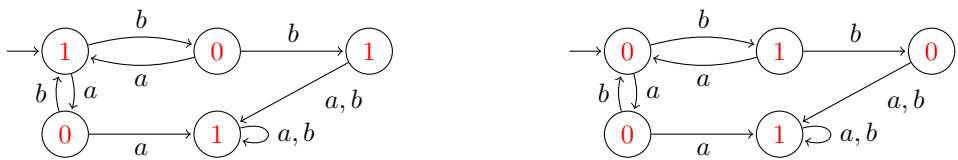
	We claim that $\cT_1\prec_2 \cT_2$ but $\cT_2\not\prec_2 \cT_1$. Starting with the latter, observe that $\cT_2(ab)=00$, but $\cT_1(ab)=\cT_1(ba)=01$. Since $00\not\req[2]01$, we have $\cT_2\not\prec_2 \cT_1$.

	We turn to show that $\cT_1\prec_{2}\cT_2$.  Observe that for every input word of the form $x\in (ab+ba)^m$, we have $\cT_1(x)=(01)^m$, and $x\req[2] (ba)^m$. So in this case we have that $\cT_2((ba)^m)=(10)^m\req[2] (01)^m$. Next, for $x\in (ab+ba)^m\cdot bb\cdot w$ for some $w\in \sI^*$ we have $\cT_1(x)=(01)^m 01 1^{|w|}$ and $x\req[2] (ba)^m\cdot bb\cdot w$, for which $\cT_2((ba)^m\cdot bb\cdot w)=(01)^m 10 1^{|w|}\req[2]\cT_1(x)$.
	The case where $x\in (ab+ba)^m\cdot aa\cdot w$ is handled similarly. We conclude that $\cT_1\prec_{2} \cT_2$.
\end{exa}

Round simulation and round equivalence give rise to the following decision problems:
\begin{itemize}
	\item In \emph{fixed round simulation} (resp. \emph{fixed round equivalence}) we are given transducers $\cT_1,\cT_2$, an \NFA for the language $\fL$, and $k>0$ in unary, and we need to decide whether $\cT_1\prec_{k,\fL} \cT_2$ (resp. whether $\cT_1\equiv_{k,\fL}\cT_2$).
	\item In \emph{existential round simulation} (resp. \emph{existential round equivalence}) we are given transducers $\cT_1,\cT_2$ and an \NFA for the language $\fL$, and we need to decide whether there exists $k>0$ such that $\cT_1\prec_{k,\fL} \cT_2$ (resp. $\cT_1\equiv_{k,\fL}\cT_2$).
\end{itemize}
In the following we identify $\fL$ with an \NFA (or \DFA) for it, as we do not explicitly rely on its description.

We start by showing that deciding equivalence (both fixed and existential) is reducible, in polynomial time, to the respective simulation problem.
\begin{lem}
	\label{lem:reduce_equiv_to_simulation}
	Fixed (resp. existential) round equivalence is Turing reducible in polynomial time to fixed (resp. existential) round simulation.
\end{lem}
\proof
	First, we can clearly reduce fixed round equivalence to fixed round simulation: given an algorithm that decides, given $\cT_1,\cT_2$, $\fL$ and $k>0$, whether $\cT_1\prec_{k,\fL} \cT_2$, we can decide whether $\cT_1\equiv_{k,\fL} \cT_2$ by using it twice to decide whether both $\cT_1\prec_{k,\fL} \cT_2$ and $\cT_1\prec_{k,\fL} \cT_2$ hold.

	A slightly more careful examination shows that the same approach can be taken to reduce existential round equivalence to existential round simulation, using the following observation: if $\cT_1\prec_{k,\fL} \cT_2$, then for every $m\in \bbN$ it holds that $\cT_1\prec_{mk,\fL}\cT_2$. Indeed, we can simply group every $m$ rounds of length $k$ and treat them as a single round of length $mk$.

	Now, given an algorithm that decides, given $\cT_1,\cT_2$ and $\fL$, whether there exists $k>0$ such that $\cT_1\prec_{k,\fL} \cT_2$, we can decide whether $\cT_1\equiv_{k,\fL} \cT_2$ by using the algorithm twice to decide whether there exists $k_1$ such that $\cT_1\prec_{k_1,\fL} \cT_2$ and $k_2$ such that $\cT_2\prec_{k_2,\fL} \cT_1$ hold. If there are no such $k_1,k_2$, then clearly $\cT_1 \not\equiv_{k,\fL} \cT_2$. However, if there are such $k_1,k_2$, then by the observation above we have $\cT_1\equiv_{k_1k_2,\fL}\cT_2$ (we can also take $\lcm(k_1,k_2)$ instead of $k_1k_2$).
\qed

By~\cref{lem:reduce_equiv_to_simulation}, for the purpose of upper-bounds, we focus henceforth on round simulation.

\section{Deciding Fixed Round Simulation}
\label{sec:deciding_fixed_round_sim}
In this section we show decidability of fixed round simulation (and, by~\cref{lem:reduce_equiv_to_simulation}, fixed round equivalence). The tools we develop will be used in~\cref{sec:deciding_existential_round_sim} to handle the existential variant.

Let $\sI$ and $\sO$ be input and output alphabets. Consider two $\sI/\sO$ transducers $\cT_1$ and $\cT_2$, and let $\fL\subseteq \sI^*$ and $k>0$.
In order to decide whether $\cT_1\prec_{k,\fL} \cT_2$, we proceed as follows. First, we cast the problem to a problem about deterministic automata. Then, we translate rounds into letters, by working over the alphabets $\sI^k$ and $\sO^k$. We construct an \NFA, dubbed the \emph{permutation closure}, for each transducer $\cT$, that captures the behaviour of $\cT$ on words and their permutations. Intuitively, the \NFA takes as input a word $(x,y)\in (\sI^{k}\times \sO^k)^*$, guesses a round-equivalent word $x'\req x$, and verifies that $\cT(x')\req \cT(x)$. We then show that round simulation amounts to deciding the containment of these \NFAs.

We now turn to give the details of the construction of these \NFAs.

\paragraph*{The trace \bf DFA} Consider a transducer $\cT=\tup{\sI,\sO,Q,q_0,\delta,\lab}$, we define its \emph{trace \DFA} $\tr(\cT)\ =$ \linebreak $\tup{\sI\times\sO,Q\cup\{q_{\bot}\},q_0,\eta,Q}$ where for $q\in Q$ and $(\sigma,\sigma')\in \sI\times \sO$ we define $\eta(q,(\sigma,\sigma'))=\delta(q,\sigma)$ if $\cT^{q}(\sigma)=\sigma'$ and $\eta(q,(\sigma,\sigma'))=q_{\bot}$ otherwise.
$q_\bot$ is a rejecting sink.

$\tr(\cT)$ captures the behaviour of $\cT$ in that $L(\tr(\cT))=\condset{(x,y)\in (\sI\times \sO)^*\!}{\cT(x)=y}$.

\paragraph*{The permutation closure \bf NFA}
Consider an \NFA $\cN=\tup{\sI\times\sO,S,s_0,\eta,F}$, and let $k>0$.
We obtain from $\cN$ an \NFA $\perm_k(\cN)=\tup{\sI^k \times \sO^k,S, s_0, \mu, F}$
where the alphabet is $\sI^k\times \sO^k$, and the transition function $\mu$ is defined as follows. For a letter $(\alpha,\beta)\in \sI^k \times \sO^k$ and a state $s\in S$, we think of $(\alpha,\beta)$ as a word in $(\sI\times \sO)^*$. Then we have
\begin{equation}
	\label{eq:perm_transition}
	\mu(s,(\alpha,\beta))=\bigcup\condset{\eta^*(s,(\alpha',\beta'))}{\fP(\alpha')=\fP(\alpha) \wedge \fP(\beta)=\fP(\beta')}.
\end{equation}

That is, upon reading $(\alpha,\beta)$, $\perm_k(\cN)$ can move to any state $s'$ that is reachable in $\cN$ from $s$
by reading a permutation of $\alpha,\beta$ (denoted $\alpha',\beta'$).
Recall that for two words $x,x'$ we have that $x\req[k]x'$ if for every two corresponding rounds $\alpha,\alpha'$ in $x$ and $x'$ we have $\fP(\alpha)=\fP(\alpha')$.
Thus, we have the following.
\begin{obs}
	\label{obs:perm_closure_language}
	In the notations above, it holds that $L(\perm_k(\cN))=\{(x,y)\in \sI^*\times \sO^* \mid \exists x'\req[k] x, y'\req[k] y, (x',y')\in L(\cN) \wedge\ |x|=|y|=kR \text{ for some }R\in \bbN\}$.
\end{obs}
Since the transition function of $\perm_k(\cN)$ is only defined using permutations of its input letters, we have the following property, which we refer to as \emph{permutation invariance}:
\begin{obs}[Permutation invariance]
	\label{obs:perm_invariance}
	For every state $s\in S$ and letters $(\alpha,\beta), (\alpha',\beta')\!\in \sI^k \times \sO^k$, if $\fP(\alpha)=\fP(\alpha')$ and $\fP(\beta)=\fP(\beta')$ then $\mu(s,(\alpha,\beta))=\mu(s,(\alpha',\beta'))$.
\end{obs}

Given a transducer $\cT$, we apply the permutation closure to the trace \DFA of $\cT$. In order to account for the restriction given by $\fL\subseteq \sI^*$, we identify it with $\fL \subseteq \sI^*\times \sO^*$. Recall that $\fL$ denotes both a language and a corresponding \NFA (or \DFA), so what this means is that the \NFA, reading input from $\sI^*\times \sO^*$, simply ignores the second component.
\begin{lem}
	\label{lem:permutation_closure_construction}
	Consider transducers $\cT_1,\cT_2$, an \NFA $\fL$ and $k>0$. Let $\cA_1^k=\perm_k(\tr(\cT_1)\cap\fL)$ (where the intersection implies the product \NFA construction) and $\cA_2^k=\perm_k(\tr(\cT_2))$, then
	\small
	\begin{flalign*}
		L(\cA_1^k) &= \condset{(x,y)\in \sI^*\times \sO^*}{\exists x'\req[k] x,\ \cT_1(x')\req[k] y\ \wedge\ |x|=|y|=kR \text{ where }R\in \bbN \wedge x'\in \fL}, & \\
		L(\cA_2^k) &= \condset{(x,y)\in \sI^*\times \sO^*}{\exists x'\req[k] x,\ \cT_2(x')\req[k] y\ \wedge\ |x|=|y|=kR \text{ where }R\in \bbN}. &
	\end{flalign*}
\end{lem}
\proof
	Recall that $\tr(\cT)$ accepts a word $(x',y')$ iff $\cT(x')=y'$. The claim then follows from~\cref{obs:perm_closure_language}, by replacing the expression $y\req y' \wedge (x',y')\in L(\tr(\cT))$ with the equivalent expression $\cT(x')\req[k] y$.
\qed

We now reduce round simulation to the containment of permutation closure \NFAs.

\begin{lem}
	\label{lem:round_equivalence_iff_perm_containment}
	Consider transducers $\cT_1,\cT_2$, an \NFA $\fL$ and $k>0$. Let $\cA_1^k=\perm_k(\tr(\cT_1)\cap\fL)$ and $\cA_2^k=\perm_k(\tr(\cT_2))$,
	then $\cT_1\prec_{k,\fL} \cT_2$ iff $L(\cA^k_1)\subseteq L(\cA^k_2)$.
\end{lem}
\proof
	For the first direction, assume $\cT_1\prec_{k,\fL} \cT_2$, and let $(x,y)\in L(\cA^k_1)$. By~\cref{lem:permutation_closure_construction}, $x$ and $y$ are $k$-round words, and there exists a word $x'\in \fL$ such that $x\req x'$ and $\cT_1(x')\req y$. Since $\cT_1\prec_{k,\fL} \cT_2$, then applying the definition on $x'$ yields that there exists a $k$-round word $x''$ such that $x'\req x''$ and such that $\cT_1(x')\req \cT_2(x'')$. Since $\req$ is an equivalence relation, it follows that $x\req x''$ and $\cT_2(x'')\req y$, so again by~\cref{lem:permutation_closure_construction} we have $(x,y)\in L(\cA^k_2)$.

	Conversely, assume $L(\cA^k_1)\subseteq L(\cA^k_2)$, we wish to prove that for every $k$-round word $x\in \fL$ there exists a word $x'$ such that $x\req x'$ and $\cT_1(x)\req \cT_2(x')$. Let $x\in \fL$ be a $k$-round word, and let $y=\cT_1(x)$, then clearly $(x,y)\in L(\cA^k_1)\subseteq L(\cA^k_2)$ (since $x\req x$, $\cT_1(x)=y\req y$ and $x\in \fL$). By~\cref{lem:permutation_closure_construction}, there exists $x'$ such that $x\req x'$ and $\cT_2(x')\req y=\cT_1(x)$, so $\cT_2(x')\req \cT_1(x)$, thus concluding the proof.
\qed

\begin{rem}
	\label{rmk:det_A1}
	The proof of~\cref{lem:round_equivalence_iff_perm_containment} does not require taking the permutation closure of $\tr(\cT_1)\cap \fL$, and it could be simplified by using instead of $\cA^k_1$, the augmentation of $\tr(\cT_1)\cap \fL$ to $k$-round words. However, such an \NFA is not permutation invariant, which is key to our solution for existential round simulation. Since this simplification does not reduce the overall complexity, we use a uniform setting for both solutions.
\end{rem}

\cref{lem:round_equivalence_iff_perm_containment} shows that deciding fixed round equivalence amounts to deciding containment of \NFAs. By analyzing the size of the \NFAs, we obtain the following.
\begin{thm}
	\label{thm:fixed_re_PSPACE}
	Given transducers $\cT_1,\cT_2$, an \NFA $\fL$, and $k>0$ in unary, the problem of deciding whether $\cT_1\prec_{k,\fL}\cT_2$ is in \PSPACE.
\end{thm}
\proof
	Let $\cA_1^k=\perm_k(\tr(\cT_1)\cap \fL)$ and $\cA_2^k=\perm_k(\tr(\cT_2))$. By~\cref{lem:round_equivalence_iff_perm_containment}, deciding whether $\cT_1\prec_{k,\fL} \cT_2$ amounts to deciding whether $L(\cA^k_1)\subseteq L(\cA^k_2)$. Looking at the dual problem, recall that for two \NFAs $\cN_1, \cN_2$ we have that $L(\cN_1)\not\subseteq L(\cN_2)$ iff
	there exists $w\in L(\cN_2)\setminus L(\cN_1)$ with $|w|\le |\cN_1|\cdot 2^{|\cN_2|}$ (this follows immediately by bounding the size of an \NFA for $L(\cN_1)\cap \overline{L(\cN_2)}$). Thus, we can decide whether $L(\cA^k_1)\subseteq L(\cA^k_2)$ by guessing a word $w$ over $\sI^k\times \sO^k$ of single-exponential length (in the size of $\cA^k_1$ and $\cA^k_2$), and verifying that it is accepted by $\cA^k_1$ and not by $\cA^k_2$.

	Observe that to this end, we do not explicitly construct $\cA^k_1$ nor $\cA^k_2$, as their alphabet size is exponential. Rather, we evaluate them on each letter of $w$ based on their construction from $\cT$. At each step we keep track of a counter for the length of $w$, a state of $\cA^k_1$, and a set of states of $\cA^k_2$. Since the number of states in $\cA^k_1$ and $\cA^k_2$ is the same as that of $\cT_1$ and $\cT_2$, this requires polynomial space.

	By Savitch's theorem we have that $\coNPSPACE=\PSPACE$, and the proof is concluded.
\qed

We now establish a \PSPACE-hardness lower bound, thus concluding that the problem is \PSPACE-complete. In fact, we show a lower bound for round equivalence. Note that a priori, this does not entail a lower bound for round simulation by~\cref{lem:reduce_equiv_to_simulation}, since the reduction there is a Turing reduction. However, our \PSPACE-hardness proof actually explicitly shows the hardness of both simulation and equivalence.
\begin{thm}
	\label{thm:equivalence_PSPACE-H}
	The problem of deciding, given transducers $\cT_1,\cT_2$, whether $\cT_1\equiv_{k,\fL} \cT_2$, is \PSPACE-hard, even for $k=2$ and $\fL$ of constant size (given as a 4-state \DFA).
\end{thm}
\proof[Proof sketch]
	We show a reduction from the universality problem for \NFAs over alphabet $\{0,1\}$ where all states are accepting and the degree of nondeterminism is at most 2. See~\cref{chap:apx} for a proof of \PSPACE-hardness of this problem and for the full reduction.

	Consider an \NFA $\cN=\tup{Q,\{0,1\},\delta,q_0,Q}$ where $|\delta(q,\sigma)|\le 2$ for every $q\in Q$ and $\sigma\in \{0,1\}$. Set $\fL=(ab+cd)^*$. We construct two transducers $\cT_1$ and $\cT_2$ over input and output alphabets $\sI=\{a,b,c,d\}$ and $\sO=\{\top,\bot\}$ such that $L(\cN)=\{0,1\}^*$ iff $\cT_1\equiv_{2,\fL}\cT_2$.

	Intuitively, our reduction encodes $\{0,1\}$ over $\{a,b,c,d\}$ by identifying $0$ with $ab$ and with $ba$, and $1$ with $cd$ and with $dc$. Then, $\cT_1$ keeps outputting $\top$ for all inputs in $\fL$, thus mimicking a universal language in $\{0,1\}^*$ (see~\cref{fig:PSPACE_reduction_T1}), whereas $\cT_2$ is obtained by replacing every nondeterministic transition of $\cN$ on e.g. 0 by two deterministic branches, on e.g. $ab$ and $ba$ (see~\cref{fig:PSPACE_reduction_T2}). Hence, when we are allowed to permute $ab$ and $ba$ by round equivalence, we capture the nondeterminism of $\cN$.

	We show that $L(\cN)=\{0,1\}^*$ iff $\cT_1\equiv_{2,\fL}\cT_2$ by showing that permuting a word $w\in \fL$ essentially amounts to choosing an accepting run of $\cN$ on the corresponding word in $\{0,1\}^*$.
\qed

\begin{cor}
	\label{cor:PSPACE-C}
	Given transducers $\cT_1,\cT_2$, an \NFA $\fL$, and $k>0$ in unary, the problem of deciding whether $\cT_1\prec_{k,\fL}\cT_2$ is \PSPACE-complete.
\end{cor}

\section{Deciding Existential Round Simulation}
\label{sec:deciding_existential_round_sim}
In~\cref{sec:deciding_fixed_round_sim}, we established a method for deciding $k$-round simulation for a given $k$. This case is for when the systems in question exhibit an apparent symmetry with a round length that a developer can guess; such as Round Robin where the round length is the number of processes involved. However, $k$ is not necessarily given in the general sense.

We turn to solve existential round simulation. That is, given $\cT_1,\cT_2$ and $\fL$, we wish to decide whether there exists $k>0$ such that $\cT_1\prec_{k,\fL} \cT_2$.
By~\cref{lem:round_equivalence_iff_perm_containment}, this is equivalent to deciding whether there exists $k>0$ such that $L(\cA^k_1)\subseteq L(\cA^k_2)$, as defined therein.

Recall that solving the decision problems of round simulation will aid us in solving the initial problem of round symmetry, which gave the motivation for this work. The transition between the problems is explained in~\cref{sec:application}.

\subsection{Intuitive Overview}
\label{sec:intuitive_overview}
We start with an intuitive explanation of the solution and its challenges. For simplicity, assume for now $\fL=\sI^*$, so it can be ignored. The overall approach is to present a practical method for hunting $k$: in~\cref{thm:bound_on_k}, the main result of this section, we give an upper bound on the minimal $k>0$ for which $\cT_1\prec_{k} \cT_2$, rendering the search space finite. In order to obtain this bound, we proceed as follows. Observe that for a transducer $\cT$ and for $0<k\neq k'$ the corresponding permutation closure \NFAs $\perm_k(\tr(\cT))$ and $\perm_{k'}(\tr(\cT))$ are defined on the same state space, but differ by their alphabet ($\sI^{k}\times \sO^{k}$ vs $\sI^{k'}\times \sO^{k'}$). Thus, by definition, these \NFAs obtained from an increasing round length form infinitely many distinct automata. Nonetheless, there are only finitely many possible types of letters (indeed, at most $|\bbB^{Q\times Q}|=2^{|Q|^2}$). Therefore, there are only finitely many \emph{type~profiles} for \NFAs -- that is, the set of letter types occurring in the \NFA -- up to multiplicities of the letter types.

Recall that by~\cref{lem:round_equivalence_iff_perm_containment}, we have $\cT_1\prec_k \cT_2$ iff $L(\perm_k(\tr(\cT_1)))\subseteq L(\perm_{k}(\tr(\cT_2)))$.
Intuitively, one could hope that if $\perm_k(\tr(\cT_i))$ and $\perm_{k'}(\tr(\cT_i))$ have the same type profile, for each $i\in \{1,2\}$, then it holds that $L(\perm_k(\tr(\cT_1)))\subseteq L(\perm_{k}(\tr(\cT_2)))$ iff $L(\perm_{k'}(\tr(\cT_1)))\subseteq L(\perm_{k'}(\tr(\cT_2)))$. Then, if one can bound the index $k$ after which no further type profiles are encountered, then the problem reduces to checking a finite number of containments.

Unfortunately, this is not the case, the reason being
that the mapping of letters induced by the equal type profiles $\perm_k(\tr(\cT_1))$ and $\perm_{k'}(\tr(\cT_1))$ may differ from the mapping induced by $\perm_k(\tr(\cT_2))$ and $\perm_{k'}(\tr(\cT_2))$, and thus one cannot translate language containment between the two pairs. We overcome this difficulty, however, by working from the start with product automata that capture the structure of both $\cT_1$ and $\cT_2$ simultaneously, and thus unify the letter mapping. We dub them \emph{redundant product automata} for their apparent redundancy.

We are now left with the problem of bounding the minimal $k$ after which no new type profiles appear.
In order to provide this bound, we show that for every type profile, the set of indices in which it occurs is semilinear. Then, by finding a bound for each type profile, we obtain the overall bound.
The main result of this section is the following.
\begin{thm}
	\label{thm:bound_on_k}
	Given transducers $\cT_1,\cT_2$ and $\fL$, we can effectively compute $K_0>0$ such that if $\cT_1\prec_{k,\fL} \cT_2$ for some $k\in \bbN$, then $\cT_1\prec_{k',\fL} \cT_2$ for some $k'\le K_0$.
\end{thm}
Which by~\cref{lem:round_equivalence_iff_perm_containment} immediately entails the following.
\begin{cor}
	\label{cor:exist_k_decidable}
	Existential round simulation is decidable.
\end{cor}

We prove~\cref{thm:bound_on_k} in~\cref{sec:proof_of_bound}, organized as follows. We start by lifting the definition of types in an \NFA to Parikh vectors, and show how these relate to the \NFA (in \cref{lem:type_of_parikh}). We then introduce Presburger arithmetic and its relation to Parikh's theorem. In~\cref{lem:parikh_type_definable} we show that the set of Parikh vectors that share a type $\tau$ is definable in Presburger arithmetic, which provides the first main step towards our bound.

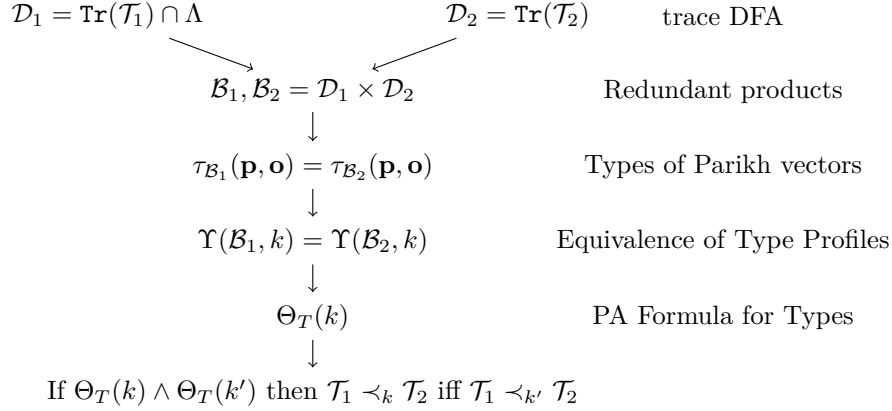
\begin{figure}[ht]
	\centering
	\begin{tikzpicture}[shorten >=1pt,node distance=1cm and 2.7cm,on grid,auto]
		\node (text_tr) [align=center] {trace \DFA};
		\node (text_redundant) [align=center, below=of text_tr] {Redundant products};
		\node (text_types) [align=center, below=of text_redundant] {Types of Parikh vectors};
		\node (text_typeprof) [align=center, below=of text_types] {Equivalence of Type Profiles};
		\node (text_Theta) [align=center, below=of text_typeprof] {PA Formula for Types};

		\node (tr2) [align=center, left=of text_tr] {$\cD_2=\tr(\cT_2)$};
		\node (redundant) [align=center, below left=of tr2] {$\cB_1, \cB_2 = \cD_1\times\cD_2$};
		\node (tr1) [align=center, above left=of redundant] {$\cD_1=\tr(\cT_1)\cap\fL$};
		\node (types) [below=of redundant, align=center] {$\tau_{\cB_1}(\mathbf{p},\mathbf{o})=\tau_{\cB_2}(\mathbf{p},\mathbf{o})$};
		\node (typeprof) [below=of types, align=center] {$\Upsilon(\cB_1, k)=\Upsilon(\cB_2, k)$};
		\node (Theta) [below=of typeprof, align=center] {$\Theta_T(k)$};
		\node (mapping) [below=of Theta, align=center] {If $\Theta_T(k)\wedge \Theta_T(k')$ then $\cT_1\prec_k\cT_2$ iff $\cT_1\prec_{k'}\cT_2$};

		\path[->]
		(tr1) edge node[] {} (redundant)
		(tr2) edge node[] {} (redundant)
		(redundant) edge node[] {} (types)
		(types) edge node[] {} (typeprof)
		(typeprof) edge node[] {} (Theta)
		(Theta)  edge node[] {} (mapping);

	\end{tikzpicture}
	\caption{A flow diagram for the proof steps in~\cref{sec:proof_of_bound}.}
	\label{fig:flow}
\end{figure}

We then proceed to define the redundant product automata mentioned above, which serve to unify the types between $\cT_1$ and $\cT_2$. In~\cref{obs:redundant_product,obs:redundant_product_for_perm_automata} we formalize the connection of these products to the transducers $\cT_1$ and $\cT_2$. Then, we formally define the type profiles and prove in~\cref{lem:type_profile_bound} that they exhibit a semilinear behaviour. Finally, in~\cref{lem:profile_equality_to_simulation} we prove that when two redundant product automata have the same type profile, then the containment mentioned above can be shown. Combining these results, we obtain~\cref{thm:bound_on_k}.
A flow diagram for the proof is illustrated in~\cref{fig:flow}.

\subsection{Proof of Theorem~\ref{thm:bound_on_k}}
\label{sec:proof_of_bound}
\paragraph*{Type matrices of Parikh vectors.}
Consider the alphabet $\sI^k\times \sO^k$ for some $k>0$.
Recall that by~\cref{obs:perm_invariance}, permutation closure \NFAs are permutation invariant, and from~\cref{sec:preliminaries}, the type of a word in an \NFA is the transition matrix it induces.
In particular, for permutation invariant \NFAs, two letters $(\alpha,\beta),(\alpha',\beta')\in \sI^k\times \sO^k$ with $\fP(\alpha)=\fP(\alpha')$ and $\fP(\beta)=\fP(\beta')$ have the same type.

Following this, we now lift the definition of types to Parikh vectors. Consider an \NFA $\cN=\tup{\sI\times \sO,S,s_0,\eta,F}$, and let $\vec{p}\in \bbN^{\sI},\vec{o}\in \bbN^{\sO}$ be Parikh vectors with $|\vec{p}|=|\vec{o}|=k$. We define the type $\tau_\cN(\vec{p},\vec{o})\in \bbB^{S\times S}$ to be $\tau_{\perm_k(\cN)}(\alpha,\beta)$ where $(\alpha,\beta)\in \sI^k\times \sO^k$ are such that $\fP(\alpha)=\vec{p}$ and $\fP(\beta)=\vec{o}$. By permutation invariance, this is well-defined, i.e. is independent of the choice of $\alpha$ and $\beta$.

Note that we use different automata to extract the type of words of different lengths. We obtain a more uniform description as follows.
\begin{lem}
	\label{lem:type_of_parikh}
	In the notations above, for every $s_1,s_2\in S$, we have $(\tau_{\cN}(\vec{p},\vec{o}))_{s_1,s_2}=1$ iff there exists $(\alpha,\beta)\in \sI^k\times \sO^k$ with $\fP(\alpha)=\vec{p}$ and $\fP(\beta)=\vec{o}$ such that $s_1\runs{(\alpha,\beta)}{\perm_k(\cN)}s_2$.
\end{lem}
\proof
	By the definitions preceding the lemma, we have that $\tau_{\cN}(\vec{p},\vec{o})=\tau_{\perm_k(\cN)}(\alpha',\beta')$ for some $(\alpha',\beta')\in \sI^k\times \sO^k$ are such that $\fP(\alpha')=\vec{p}$ and $\fP(\beta')=\vec{o}$. According to the transition function of $\perm_k(\cN)$ (as defined in~\cref{sec:deciding_fixed_round_sim}), for every $s_1,s_2\in S$ we have that $s_1\runs{(\alpha',\beta')}{\perm_k(\cN)}s_2$ iff there exist $(\alpha,\beta)\in \sI^k\times \sO^k$ with $\fP(\alpha)=\fP(\alpha')=\vec{p}$ and $\fP(\beta)=\fP(\beta')=\vec{o}$ such that $s_1\runs{(\alpha,\beta)}{\cN}s_2$. Since the type encodes the reachable pairs of states, this concludes the proof.
\qed

\paragraph*{Presburger arithmetic.}
The first ingredient in the proof of~\cref{thm:bound_on_k} is to characterize the set of Parikh vectors whose type is some fixed matrix $\tau\in \bbB^{Q\times Q}$. For this characterization, we employ the first-order theory of the naturals with addition and order $\textrm{Th}(\bbN,0,1,+,<,=)$, commonly known as \emph{Presburger arithmetic (PA)}. We do not give a full exposition of \PA but refer the reader to~\cite{Haase2018} (and references therein) for a survey. In the following we briefly cite the results we need.

For our purposes, a \PA~formula $\varphi(x_1,\ldots,x_d)$, where $x_1,\ldots, x_d$ are free variables, is evaluated over $\bbN^d$, and \emph{defines} the set $\condset{(a_1,\ldots,a_d)\in \bbN^d}{(a_1,\ldots,a_d)\models \varphi(x_1,\ldots,x_d)}$. For example, the formula $\varphi(x_1,x_2):=x_1< x_2\wedge \exists y.\ x_1=2y$ defines the set $\left\{ \,(a,b)\in \bbN^2\ \middle| \right.$ $\left. a<b\wedge a \text{ is even}\, \right\}$. 

A fundamental result about \PA is that the definable sets in \PA are exactly the semilinear sets. In particular, Parikh's theorem states that for every \NFA $\cA$, $\fP(L(\cA))$ is \PA~definable. In fact, by~\cite{Verma2005}, one can efficiently construct a linear-sized existential \PA~formula for $\fP(L(\cA))$.
We can now show that the set of Parikh vectors whose type is $\tau$ is \PA~definable.

\begin{lem}
	\label{lem:parikh_type_definable}
	Consider an \NFA $\cN=\tup{\sI\times \sO,S,s_0,\eta,F}$, and a type $\tau\in \bbB^{S\times S}$, then the set $\condset{(\vec{p},\vec{o})\in \bbN^{\sI} \times \bbN^{\sO}}{\tau_{\cN}(\vec{p},\vec{o})=\tau}$ is \PA~definable.
\end{lem}
\proof
	Let $\tau\in \bbB^{S\times S}$, and consider a Parikh vector $(\vec{p},\vec{o})\in \bbN^{\sI} \times \bbN^{\sO}$ with $k=|\vec{p}|=|\vec{o}|$. By~\cref{lem:type_of_parikh}, we have that $\tau_{\cN}(\vec{p},\vec{o})=\tau$ iff the following holds for every $s_1,s_2\in S$: we have $\tau_{s_1,s_2}=1$ iff
	there exists a letter $(\alpha,\beta)\in \sI^k\times \sO^k$ such that $\fP(\alpha)=\vec{p},\fP(\beta)=\vec{o}$, and $s_1\runs{(\alpha,\beta)}{\cN} s_2$.

	Consider $s_1,s_2\in S$ and define $\cN^{s_1}_{s_2}$ to be the \NFA obtained from $\cN$ by setting the initial state to be $s_1$ and a single accepting state $s_2$.
	Then, we have $s_1\runs{(\alpha,\beta)}{\cN} s_2$ iff $(\alpha,\beta)\in L(\cN^{s_1}_{s_2})$.

	Thus, $\tau_{\cN}(\vec{p},\vec{o})=\tau$ iff for every $s_1,s_2\in S$ we have that $\tau_{s_1,s_2}=1$ iff there exists a word $(\alpha,\beta)$ with $\fP(\alpha')=\vec{p}$ and $\fP(\beta')=\vec{o}$ such that $(\alpha,\beta)\in L(\cN^{s_1}_{s_2})$.
	Equivalently, we have $\tau_{\cN}(\vec{p},\vec{o})=\tau$ iff for every $s_1,s_2\in S$ it holds that $\tau_{s_1,s_2}=1$ iff $(\vec{p},\vec{o})\in \fP(L(\cN^{s_1}_{s_2}))$.

	By Parikh's theorem, for every $s_1,s_2\in S$ we can compute a \PA~formula $\psi_{s_1,s_2}$ such that $(\vec{p},\vec{o})\models \psi_{s_1,s_2}$ iff $(\vec{p},\vec{o})\in \fP(L(\cN^{s_1}_{s_2}))$. Now we can construct a \PA~formula $\Psi_{\tau}$ such that $\tau_{\cN}(\vec{p},\vec{o})=\tau$ iff $(\vec{p},\vec{o})\models \Psi_\tau$, as follows:
	\[\Psi_\tau:= \bigwedge_{s_1,s_2\ST \tau_{s_1,s_2}=1} \psi_{s_1,s_2}\wedge \bigwedge_{s_1,s_2\ST \tau_{s_1,s_2}=0}\neg \psi_{s_1,s_2}.\]

	Finally, observe that $\Psi_\tau$ defines the set in the premise of the lemma, so we are done.
\qed

\paragraph*{The redundant product construction.}
As mentioned in~\cref{sec:intuitive_overview}, for the remainder of the proof we want to reason about the types of $\perm_k(\tr(\cT_1)\cap \fL)$ and $\perm_k(\tr(\cT_2))$ simultaneously. In order to do so, we present an auxiliary product construction.

Let $\cT_1,\cT_2$ be transducers, $\fL\subseteq \sI^*$ be given by an \NFA, and let $\cD_1=\tr(\cT_1)\cap \fL$ and $\cD_2=\tr(\cT_2)$.
We now consider the product automaton of $\cD_1$ and $\cD_2$, and endow it with two different acceptance conditions, capturing that of $\cD_1$ and $\cD_2$, respectively. Formally, for $i\in \{1,2\}$, denote $\cD_i=\tup{\sI\times\sO,S_i,s^i_0,\eta_i,F_i}$, then the product automaton is defined as $\cB_i=\tup{\sI\times\sO,S_1\times S_2,(s^1_0,s^2_0),\eta_1\times \eta_2,G_i}$, where $G_1=F_1\times Q_2$ and $G_2=Q_1\times F_2$, and $\eta_1\times \eta_2$ denotes the standard product transition function, namely $\eta_1\times\eta_2((s_1,s_2),(\sigma,\sigma'))=(\eta_1(s_1,(\sigma,\sigma')),\eta_2(s_2,(\sigma,\sigma')))$. Thus, $\cB_i$ tracks both $\cD_1$ and $\cD_2$, but has the same acceptance condition as $\cD_i$. This seemingly ``redundant'' product construction has the following important properties, which are crucial for our proof:
\begin{obs}
	\label{obs:redundant_product}
	In the notations above, we have the following:
	\begin{enumerate}
		\item $L(\cB_1)=L(\cD_1)$ and $L(\cB_2)=L(\cD_2)$.
		\item For every letter $(\sigma,\sigma')\in \sI\times\sO$, we have $\tau_{\cB_1}(\sigma,\sigma')=\tau_{\cB_2}(\sigma,\sigma')$.
	\end{enumerate}
\end{obs}

Indeed, Item $1$ follows directly from the acceptance condition, and Item $2$ is due to the identical transition function of $\cB_1$ and $\cB_2$.

By~\cref{obs:perm_closure_language}, $L(\perm_k(\cD_i))$ depends only on $L(\cD_i)$. We thus have the following.
\begin{obs}
	\label{obs:redundant_product_for_perm_automata}
	The following holds for every $k>0$:
	\begin{enumerate}
		\item $L(\perm_k(\cB_1))=L(\perm_k(\tr(\cT_1)\cap \fL))$.
		\item $L(\perm_k(\cB_2))=L(\perm_k(\tr(\cT_2)))$.
	\end{enumerate}
\end{obs}

\paragraph*{Type profiles.}
We now consider the set of types induced by the redundant product automata $\cB_1$ and $\cB_2$ on Parikh vectors of words of length $k$. By Item 2 of~\cref{obs:redundant_product}, it is enough to consider $\cB_1$.

For $k>0$, we define the \emph{$k$-th type profile} of $\cB_1$ to be the set of all types of Parikh vectors $(\vec{p},\vec{o})$ with $|\vec{p}|=|\vec{o}|=k$ that are induced by $\cB_1$; i.e. it is the set $\Upsilon(\cB_1,k)=\condset{\tau_{\cB_1}(\fP(\alpha),\fP(\beta))}{(\alpha,\beta)\in \sI^k \times \sO^k}$. Clearly, there is only a finite number of type profiles, as $\Upsilon(\cB_1,k)\subseteq \bbB^{S'\times S'}$, where $S'$ is the state space of $\cB_1$. Therefore, as $k$ increases, after some finite $K_0$, every type profile that is ever attained will have been encountered already. We now place an upper bound on $K_0$.

\begin{lem}
	\label{lem:type_profile_bound}
	We can effectively compute $K_0>0$ such that for every $k>0$ there exists $k'\le K_0$ with $\Upsilon(\cB_1,k')=\Upsilon(\cB_1,k)$.
\end{lem}
\proof
	Consider a type $\tau$, and let $\Psi_\tau$ be the \PA~formula constructed as per~\cref{lem:parikh_type_definable} for the \NFA $\cB_1$. Observe that for a Parikh vector $(\vec{p},\vec{o})$ and for $k>0$, the expression $|\vec{p}|=|\vec{o}|=k$ is \PA definable. Indeed, writing $\vec{p}=(x_1,\ldots,x_{|\sI|})$ and $\vec{q}=(y_1,\ldots,y_{|\sO|})$, the expression is defined by $x_1+\ldots+x_{|\sI|}=k \wedge y_1+\ldots+y_{|\sO|}=k$.

	Let $T\subseteq \bbB^{S'\times S'}$ be a set of types (i.e., a potential type profile). We define a \PA~formula $\Theta_T(z)$ over a single free variable $z$ such that $k\models \Theta_T(z)$ iff $\Upsilon(\cB_1,k)=T$, as follows.
	\begin{align*}
		\Theta_T(z)&=\left(\forall \vec{p},\vec{o}, |\vec{p}|=|\vec{o}|=z \to \bigvee_{\tau\in T}\Psi_\tau(\vec{p},\vec{o})\right)
		\wedge \left(\bigwedge_{\tau\in T}\exists \vec{p},\vec{o}, |\vec{p}|=|\vec{o}|=z \wedge \Psi_\tau(\vec{p},\vec{o})\right)
	\end{align*}
	Intuitively, $\Theta_T(z)$ states that every Parikh vector $(\vec{p},\vec{o})$ with $|\vec{p}|=|\vec{o}|=z$ has a type within $T$, and that all the types in $T$ are attained by some such Parikh vector.

	By~\cite{Fischer1974,Borosh1976}, we can effectively determine for every $T$ whether $\Theta_T(z)$ is satisfiable and, if it is, find a witness $M_T$ such that $M_T\models \Theta_T(z)$. By doing so for every set $T\subseteq \bbB^{S'\times S'}$, we can set $K_0=\max\condset{M_T}{\Theta_T(z) \text{ is satisfiable}}$. Then, for every $k>K_0$ if $\Upsilon(\cB_1,k)=T$, then $T$ has already been encountered at $M_T\le K_0$, as required.
\qed

The purpose of the bound $K_0$ obtained in~\cref{lem:type_profile_bound} is to bound the minimal $k$ for which $\cT_1\prec_{k,\fL} \cT_2$, or equivalently $L(\perm_{k}(\cB_1))\subseteq L(\perm_{k}(\cB_2))$ (by~\cref{lem:round_equivalence_iff_perm_containment,obs:redundant_product_for_perm_automata}).
This is captured in the following.

\begin{lem}
	\label{lem:profile_equality_to_simulation}
	Let $k,k'>0$ such that $k\neq k'$ and $\Upsilon(\cB_1,k')=\Upsilon(\cB_1,k)$, then we have
	$L(\perm_{k}(\cB_1))\subseteq L(\perm_{k}(\cB_2))$ iff $L(\perm_{k'}(\cB_1))\subseteq L(\perm_{k'}(\cB_2))$.
\end{lem}
\proof

	By the symmetry between $k$ and $k'$, it suffices to prove w.l.o.g. that if $L(\perm_{k}(\cB_1))\subseteq\ L(\perm_{k}(\cB_2))$, then $L(\perm_{k'}(\cB_1))\subseteq L(\perm_{k'}(\cB_2))$.

	Assume the former, and let $w=(x',y')\in L(\perm_{k'}(\cB_1))$, where $(x',y')\in (\sI^{k'}\times \sO^{k'})^*$, and we denote $(x',y')=(\alpha'_1,\beta'_1)\cdots (\alpha'_n,\beta'_n)$ with $(\alpha'_j,\beta'_j)\in \sI^{k'}\times \sO^{k'}$ for every $1\le j\le n$.

	Since $\Upsilon(\cB_1,k')=\Upsilon(\cB_1,k)$, there is a mapping $\varphi$ that takes every letter $(\alpha_j',\beta_j')\in \sI^{k'}\times \sO^{k'}$ in $w$ to a letter $(\alpha_j,\beta_j)\in\sI^{k}\times \sO^{k}$ that has same type in $\perm_{k}(\cB_1)$, so that we can find $(x,y)=(\alpha_1,\beta_1)\cdots (\alpha_n,\beta_n)$ such that for every $1\le j\le n$ we have $\tau_{\cB_1}(\fP(\alpha_j),\fP(\beta_j))=\tau_{\cB_1}(\fP(\alpha'_j),\fP(\beta'_j))$.

	By the definition of the type of a Parikh vector, we have that \[\tau_{\perm_k(\cB_1)}(\alpha_j,\beta_j)=\tau_{\cB_1}(\fP(\alpha_j),\fP(\beta_j))=\tau_{\cB_1}(\fP(\alpha'_j),\fP(\beta'_j))=\tau_{\perm_{k'}(\cB_1)}(\alpha'_j,\beta'_j).\]
	In particular, since the type of a word is the concatenation (i.e., Boolean matrix product) of its underlying letters, we have that $\tau_{\perm_k(\cB_1)}(x,y)=\tau_{\perm_{k'}(\cB_1)}(x',y')$. Since $(x',y')\in L(\perm_{k'}(\cB_1))$, it follows that also $(x,y)\in L(\perm_{k}(\cB_1))$. Indeed,
	$(\tau_{\perm_{k'}(\cB_1)}(x',y'))_{s^1_0,s^1_f}=1$ where $s^1_0$ and $s^1_f$ are an initial state and an accepting state of $\perm_{k'}(\cB_1)$, respectively. But the equality of types implies $\left(\tau_{\perm_{k}(\cB_1)}(x,y)\right)_{s^1_0,s^1_f}=1$ as well, so $\perm_k(\cB_1)$ has an accepting run on $(x,y)$.

	By our assumption, $L(\perm_{k}(\cB_1))\subseteq L(\perm_{k}(\cB_2))$, so $(x,y)=\varphi(w)\in L(\perm_{k}(\cB_2))$, or equivalently, $\varphi(w)\in L(\perm_{k}(\cB_2))$.
	We now essentially reverse the arguments above, but with $\cB_2$ instead of $\cB_1$. However, this needs to be done carefully, so that the mapping of letters lands us back at $(x',y')$, and not a different word.
	Thus, instead of finding a round equivalent word, we observe that for every $1\le j\le n$, we also have
	\[\tau_{\perm_k(\cB_2)}(\alpha_j,\beta_j)=\tau_{\cB_2}(\fP(\alpha_j),\fP(\beta_j))=\tau_{\cB_2}(\fP(\alpha'_j),\fP(\beta'_j))=\tau_{\perm_{k'}(\cB_2)}(\alpha'_j,\beta'_j),\]
	This follows from Item 2 in~\cref{obs:redundant_product} and the fact that the permutation closure depends only on the transitions (and not on accepting states, which are the only difference between $\cB_1$ and $\cB_2$).

	Thus, similarly to the arguments above, we have that $(x',y')\in L(\perm_{k'}(\cB_2))$, and the mapping applied is in fact the the inverse map $\varphi^{-1}$, where $\varphi^{-1}(\varphi(w))=w$. We conclude that $L(\perm_{k'}(\cB_1))\subseteq L(\perm_{k'}(\cB_2))$, as required.

\begin{figure}[ht]
	\centering
	\begin{tikzpicture}[shorten >=1pt,node distance=2.4cm and 8cm,on grid,auto]
		\node (perm_k_1) [align=center] {$w\in\perm_{k'}(\cB_1)$ \\ \footnotesize $w=(x_1,y_1)(x_2,y_2)\cdots (x_n,y_n)$};
		\node (perm_kp_1) [below=of perm_k_1, align=center] {$\varphi(w)\in\perm_{k}(\cB_1)$ \\ \footnotesize $\varphi(w)=\varphi\left((x_1,y_1)\right)\varphi\left((x_2,y_2)\right)\cdots \varphi\left((x_n,y_n)\right)$};
		\node (perm_kp_2) [right=of perm_kp_1, align=center] {$\varphi(w)\in\perm_{k}(\cB_2)$};
		\node (perm_k_2) [right=of perm_k_1, align=center] {$\varphi^{-1}(\varphi(w))=w\in\perm_{k'}(\cB_2)$};
		\path[->]
		(perm_k_1)  edge node[fill=white, anchor=center, pos=0.5] {$\varphi$} (perm_kp_1)
		(perm_kp_1)  edge node[fill=white, anchor=center, pos=0.5] {$\subseteq$} ($(perm_kp_2)+(-1.6cm,0cm)$)
		(perm_kp_2)  edge node[fill=white, anchor=center, pos=0.5] {$\varphi^{-1}$} (perm_k_2);
	\end{tikzpicture}
	\caption{A diagram for the proof structure of~\cref{lem:profile_equality_to_simulation}.}
	\label{fig:type_profile}
\end{figure}
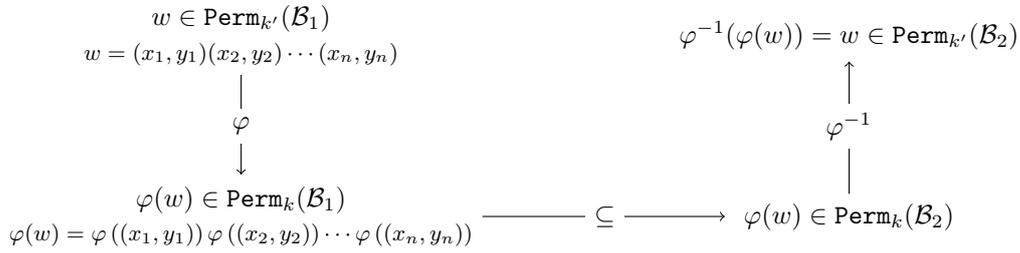

	The mapping is illustrated in~\cref{fig:type_profile}.
\qed

Combining~\cref{lem:type_profile_bound,lem:profile_equality_to_simulation}, we can effectively compute $K_0$ such that if it holds that $L(\perm_{k}(\cB_1))\subseteq L(\perm_{k}(\cB_2))$ for some $k$, then this also holds for some $k<K_0$. Finally, using~\cref{lem:round_equivalence_iff_perm_containment}, this concludes the proof of~\cref{thm:bound_on_k}. \qed

\begin{rem}[Complexity results for~\cref{thm:bound_on_k,cor:exist_k_decidable}]
	\label{rmk:complexity}
	Let $n$ be the number of states in $\cT_1\times \cT_2$. Observe that the formula $\Psi_\tau$ constructed in~\cref{lem:parikh_type_definable} comprises a conjunction of $O(n^2)$ \PA~subformulas, where each subformula is either an existential \PA~formula of length $O(n)$, or the negation of one. Then, the formula $\Theta_T$ in~\cref{lem:type_profile_bound} consists of a universal quantification, nesting a disjunction over $|T|$ formulas of the form $\Psi_\tau$, conjuncted with $|T|$ existential quantifications, nesting a single $\Psi_\tau$ each.
	Overall, this amounts to a formula of length $|T|\le 2^{n^2}$, with alternation depth 3.~\footnote{Alternation depth is usually counted with the outermost quantifier being existential, which is not the case here, hence $3$ instead of $2$.}

	Using quantifier elimination~\cite{Cooper1972,Oppen1978}, we can obtain a witness for the satisfiability of $\Theta_T$ of size 4-exponential in $n^2$. Then, finding the overall bound $K_0$ amounts to $2^{2^{n^2}}$ calls to find such witnesses. Finally, we need $K_0$ oracle calls to~\cref{lem:round_equivalence_iff_perm_containment} in order to decide existential simulation, and since $K_0$ may have a 4-exponential size description, this approach yields a whopping $\EEEEEXP$ algorithm.
	This approach, however, does not exploit any of the structure of $\Theta_T$.
\end{rem}

\subsection{Lower Bounds for Existential Round Simulation}
\label{sec:lower_bounds_existential}
The complexity bounds in~\cref{rmk:complexity} are naively analyzed, and we leave it for future work to conduct a more in-depth analysis. In this section, we present lower bounds to delimit the complexity gap. Note that there are two relevant lower bounds: one on the complexity of deciding round simulation, and the other on the minimal value of $K_0$ in~\cref{thm:bound_on_k}.

We start with the complexity lower bound, which applies already for round equivalence.

\begin{thm}
	\label{thm:existential_equivalence_PSPACE-H}
	The problem of deciding, given transducers $\cT_1,\cT_2$, whether $\cT_1\equiv_{k,\fL} \cT_2$ for \emph{any} $k$, is \PSPACE-hard, even for $\fL$ of a constant size (given as a 5-state \DFA).
\end{thm}
\proof[Proof sketch]
	We present a similar reduction to that of~\cref{thm:equivalence_PSPACE-H} from universality of \NFAs (see~\cref{apx:proof_existential_PSPACE-H}). In order to account for the unknown value of $k$, we allow padding words with a fresh symbol $\#$, which is essentially ignored by the transducers.
\qed

Next, we show that the minimal value for $K_0$ can be exponential in the size of the given transducers (in particular, of $\cT_2$).

\begin{exa}[Exponential round length]
	\label{example:exponential_round_length}

	Let $p_1,p_2,\ldots,p_m$ be the first $m$ prime numbers.
	We define two transducers $\cT_1$ and $\cT_2$ over input and output alphabet $\cP=\{1,\dots,m\}$, as depicted in~\cref{fig:prime} for $m=3$. Intuitively, $\cT_1$ reads input $w\in \fL=(1\cdot 2\cdots m)^*$ and simply outputs $w$, whereas $\cT_2$ works by reading a letter $i\in \cP$, and then outputting $i$ for $p_i$ steps (while reading $p_i$ arbitrary letters) before getting ready to read a new letter $i$.

	In order for $\cT_2$ to $k$-round simulate $\cT_1$, it must be able to output a permutation of $(1\cdot 2\cdots m)^*$. In particular, the number of $1$'s, $2$'s, etc. must be equal, so $k$ must divide every prime up to $p_m$, hence it must be exponential in the size of $\cT_2$.

	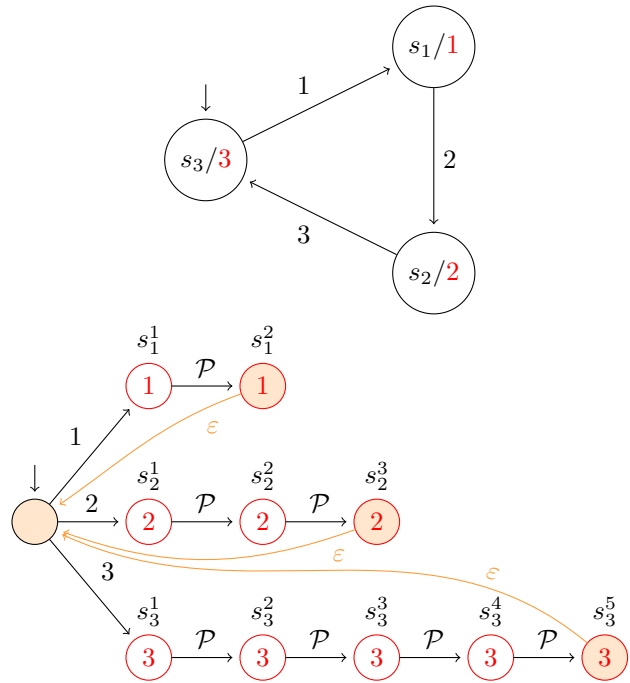
\begin{figure}[ht]
		\centering
		\begin{tikzpicture}[shorten >=1mm,node distance=1.5cm and 3cm,on grid,auto]
			\node[state, initial text={}] (q_c) [initial above] {$s_3/{\color{red}3}$};
			\node[state] (q_a) [above right=of q_0] {$s_1/{\color{red}1}$};
			\node[state] (q_b) [below right=of q_0] {$s_2/{\color{red}2}$};
			\path[->]
			(q_c) edge node {$1$} (q_a)
			(q_a) edge node {$2$} (q_b)
			(q_b) edge node {$3$} (q_c);
		\end{tikzpicture}%
	   \quad 	
		\begin{tikzpicture}[shorten >= 1mm, node distance=1.8cm and 1.5cm,on grid,auto]
			\tikzstyle{smallnode}=[circle, inner sep=0mm, outer sep=0mm, minimum size=6mm, draw=black];

			\node[smallnode, fill=orange!20, initial text={}] (q_0) [initial above] {};
			\node[smallnode, label={$s^1_2$}, red] (q_b1) [right=of q_0] {$2$};
			\node[smallnode, label={$s^2_2$}, red] (q_b2) [right=of q_b1] {$2$};
			\node[smallnode, label={$s^3_2$}, red, fill=orange!20] (q_b3) [right=of q_b2] {$2$};
			\node[smallnode, label={$s^1_1$}, red] (q_a1) [above=of q_b1] {$1$};
			\node[smallnode, label={$s^2_1$}, red, fill=orange!20] (q_a2) [right=of q_a1] {$1$};
			\node[smallnode, label={$s^1_3$}, red] (q_c1) [below=of q_b1] {$3$};
			\node[smallnode, label={$s^2_3$}, red] (q_c2) [right=of q_c1] {$3$};
			\node[smallnode, label={$s^3_3$}, red] (q_c3) [right=of q_c2] {$3$};
			\node[smallnode, label={$s^4_3$}, red] (q_c4) [right=of q_c3] {$3$};
			\node[smallnode, label={$s^5_3$}, red, fill=orange!20] (q_c5) [right=of q_c4] {$3$};
			\path[->]
			(q_0) edge node {$1$} (q_a1)
			edge node {$2$} (q_b1)
			edge node {$3$} (q_c1)
			(q_a1) edge node {$\cP$} (q_a2)
			(q_a2) edge [out=200, in=35, blue!80, pos=0.2] node {$\varepsilon$} (q_0)
			(q_b1) edge node {$\cP$} (q_b2)
			(q_b2) edge node {$\cP$} (q_b3)
			(q_b3) edge [out=200, in=-20, blue!80, pos=0.1] node {$\varepsilon$} (q_0)
			(q_c1) edge node {$\cP$} (q_c2)
			(q_c2) edge node {$\cP$} (q_c3)
			(q_c3) edge node {$\cP$} (q_c4)
			(q_c4) edge node {$\cP$} (q_c5)
			(q_c5) edge [out=140, in=-25, blue!80, pos=0.2, sloped, above] node {$\varepsilon$} (q_0);
		\end{tikzpicture}
		\caption{The transducers $\cT_1$ (left) and $\cT_2$ (right) for $m=3$ in~\cref{example:exponential_round_length}.
		The transition $s\runs{\varepsilon}{}t$ in $\cT_2$ means that the transition function from state $s$ behaves identically as from $t$.
		}
		\label{fig:prime}
	\end{figure}

	The sum of the number of states in $\cT_1$ and $\cT_2$ is $1+m+\sum_{i=1}^m p_i = \mathsf{O}\left(\sum_{i=1}^m p_i\right)$. Set $Q=\prod_{i=1}^m p_i$. It is easily verified that $\cT_1\prec_k \cT_2$ holds for $k = m\cdot Q$, which is exponential in the number of states. Indeed, for the round $w=(1\cdots m)^{Q}$, we consider the permutation $1^{Q}\cdots m^{Q}$, on which the run of $\cT_2$ induces the same output.

	We now show that this $k$ is minimal.
	For a word $x\in (1\cdot 2\cdots m)^*$ in rounds of $k$ to have round equivalent outputs in $\cT_1$ and $\cT_2$, there must be some word round equivalent word $x'$ in which every appearance of $i\in\cP$ is part of a sequence of appearances of $i$, of length $p_i$, except maybe at its end. If $m\mid k$, then there are $\frac{k}{m}$ appearances of each $i$, so $\frac{k}{m}$ must be divisible by all primes, except maybe one. The latter possibility is falsified when considering the next round. If, however, $m\nmid k$, then in the next round, $1\in\cP$ will have one less appearance than in the first round. This, again, makes impossible the round equivalence of the outputs when considering one additional round.

\end{exa}

\section{From Process Symmetry to Round Equivalence}
\label{sec:application}
As mentioned in~\cref{sec:introduction}, our original motivation for studying round simulation comes from process symmetry. We present process symmetry with an example before introducing the formal model.
Recall the Round Robin scheduler from~\cref{example:transducer-req}. There, at each time step, the scheduler receives as input the IDs of processes in $\cP=\{0,1,2\}$ that are making a request, and it responds with the IDs of those that are granted (either a singleton $\{i\}$ or $\emptyset$).

In process symmetry, we consider a setting where the identifiers of the processes may be permuted. This corresponds to the IDs representing, for instance, ports, and the processes not knowing which port they are plugged into. Thus, the input received may be a permutation of the actual identifiers of the processes. Note that a permutation in this case is a bijection over identifiers, not indices as in previous sections. Then, we say that a transducer is \emph{process symmetric} if the outputs are permuted in a way that matches the permutation of identifiers. For example, in the RR scheduler of \cref{example:transducer-req}, the output corresponding to input $\{1,2\}\{3\}\{3\}$ is $\{1\}\emptyset\{3\}$. However, if we permute the identifiers by swapping processes $1$ and $3$, we obtain the input $\{3,2\}\{1\}\{1\}$. Then, the output of RR is $\emptyset\emptyset\emptyset$, demonstrating that RR is not process symmetric. Indeed, the output letters have to be permuted in the same manner as the input for RR to be process symmetric. 

In~\cite{Almagor2020b}, several definitions of process symmetry are studied for probabilistic transducers. In the deterministic case, however, process symmetry is a very strict requirement. In order to overcome this, we allow some flexibility by letting the transducer do local reordering in the word to account for the input permutation. For instance, if we are allowed to rearrange the input $\{3,2\}\{1\}\{1\}$ to $\{1\}\{1\}\{3,2\}$, then the output becomes $\{1\}\emptyset\{3\}$, and once we apply the inverse permutation, this becomes $\{3\}\emptyset\{1\}$. This, in turn, can be again rearranged to obtain the original output $\{1\}\emptyset\{3\}$.
In this sense, the scheduler is ``locally stable'' against permutations of the identifiers of processes.

We now turn to give the formal model.
Consider a set of processes $\cP=\{1,\dots,m\}$ and $k>0$. For a permutation $\pi$ of $\cP$ (i.e. a bijection $\pi:\cP\to \cP$) and a letter $\sigma\in 2^{\cP}$, we obtain $\pi(\sigma)=\{\pi(i):i\in \sigma\}\in 2^{\cP}$ by applying $\pi$ to each process in $\sigma$. We lift this to words $x\in (2^{\cP})^*$ by applying the permutation letter-wise to obtain $\pi(x)$.
We now say that a $2^\cP/2^\cP$ transducer $\cT=\tup{2^\cP,2^\cP,Q,q_0,\delta,\lab}$ is \emph{$k$-round symmetric} if for every permutation $\pi$ of $\cP$ and for every $k$-round word $x\in (2^{\cP})^*$ there exists $x' \in (2^{\cP})^*$ such that $\pi(x)\req[k] x'$ and $\pi(\cT(x))\req[k] \cT(x')$.
We say that $\cT$ is $k$-round symmetric \WRT $\pi$ if the above holds for a fixed permutation $\pi$.

\begin{exa}
	\label{example:rr_rsym}
	Consider the RR scheduler for $n$ processes (cf.~\cref{example:transducer-req}), and let $\cT$ be a transducer for it. As discussed above, $\cT$ is not process symmetric. Intuitively, however, RR is symmetric in the sense that all processes are ``treated equally'' within each round. We now show that round symmetry captures this property.

    Consider for example the input word $x=\{0,2\}\{1\}\{2\}$ over $\cP=\{0,1,2\}$, and let $\pi=(0\ 1)$ be a permutation swapping processes $0$ and $1$. We have that $\pi(x)=\{1,2\}\{0\}\{2\}$. Observe that $\cT(x)=\{0\},\{1\},\{2\}$, meaning all processes are granted. We can now choose $x'=\{0\}\{1,2\}\{2\}$ so that $x'\req[3]x$, and we have that $\cT(x')=\{0\},\{1\},\{2\}$. and in particular $\cT(x')\req[3] \pi(\cT(x))$, since $\pi(\cT(x))=\{1\},\{0\},\{2\}$.

    In general, consider a permutation $\pi\in\cS_n$ applied to the signals. We can then preserve the behaviour of the system (i.e. the identifiers of the process that receive grants) by reordering the requests. Indeed, given input $x$, consider the $i$-th round $b_1 b_2\cdots b_n$ of $\pi(x)$. We obtain $x'$ by setting the $i$-th round to $b_{\pi^{-1}(1)} b_{\pi^{-1}(2)} \cdots b_{\pi^{-1}(n)}$.
	Then, it holds that $\cT(x)=\pi^{-1}(\cT(x'))$ or equivalently, $\pi(\cT(x))=\cT(x')$, so RR is $n$-round symmetric.
\end{exa}

\Cref{example:rr_rsym} shows that RR exhibits round symmetry \WRT all permutations. In the general sense, round symmetry might hold \WRT some permutations but not others, as is the case in the following.

\begin{exa}
	\label{example:rsym}
	Fix $\cP=\{0,1,2\}$ and let $\cT$ be the $2^\cP/2^\cP$ transducer illustrated in~\cref{fig:example_rsym}. It is not difficult to see that $\cT$ satisfies 2-round symmetry \WRT $\pi=(0\ 1)$ but not \WRT  e.g. $(0\ 2)$.

	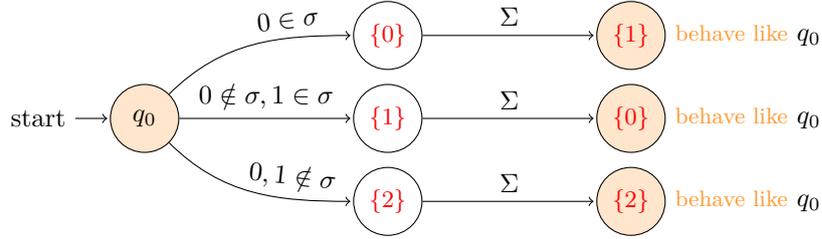
\begin{figure}[ht]
		\centering
		\begin{tikzpicture}[shorten >=1pt,node distance=1.1cm and 3.2cm,on grid,auto]
			\tikzset{every state/.style={minimum size=9mm, inner sep=0}};
			\node[state] (q_0)  [initial, fill=orange!20] {$q_0$};
			\node[state] (s_0)  [above right=of q_0, text=red] {\small $\{0\}$};
			\node[state] (s_1)  [right=of q_0, text=red] {\small $\{1\}$};
			\node[state] (s_2)  [below=of s_1, text=red] {\small $\{2\}$};
			\node[state] (p_0)  [right=of s_0, text=red, fill=orange!20, label={right:{\footnotesize\color{blue!80} behave like} $q_0$}] {\small $\{1\}$};
			\node[state] (p_1)  [right=of s_1, text=red, fill=orange!20, label={right:{\footnotesize\color{blue!80} behave like} $q_0$}] {\small $\{0\}$};
			\node[state] (p_2)  [right=of s_2, text=red, fill=orange!20, label={right:{\footnotesize\color{blue!80} behave like} $q_0$}] {\small $\{2\}$};

			\path[->]
			(q_0) edge [sloped, pos=0.7, out=45, in=180] node {$0\in\sigma$} (s_0)
			(q_0) edge [sloped, pos=0.5] node {$0\notin\sigma, 1\in\sigma$} (s_1)
			(q_0) edge [sloped, pos=0.7, out=-45, in=180] node {$0,1\notin\sigma$} (s_2)
			(s_0) edge [] node {$\Sigma$} (p_0)
			(s_1) edge [] node {$\Sigma$} (p_1)
			(s_2) edge [] node {$\Sigma$} (p_2);
		\end{tikzpicture}
		\caption{Transducer $\cT$ satisfying round symmetry \WRT $\pi=(0\ 1)$ but not $(0\ 2)$.}
		\label{fig:example_rsym}
	\end{figure}
\end{exa}

The central decision problems in round symmetry are akin to those of round simulation: in \emph{fixed round symmetry} we are given $\cT$ and $k$ and we ask whether $\cT$ is $k$-round symmetric, and in \emph{existential round symmetry} we ask whether there exists $k>0$ for which this holds. 
Observe that for round symmetry we have $\fL=(2^{\cP})^*$, and is therefore ignored in the following.

\paragraph*{From round symmetry to round simulation.}
\label{sec:symmetry_to_simulation}
As we now show, round symmetry can be cast to the setting of round simulation.
We start with the case where the permutation $\pi$ is given.

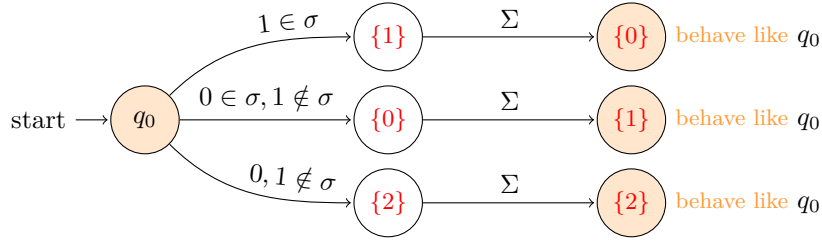
\begin{figure}[ht]
	\centering
	\begin{tikzpicture}[shorten >=1pt,node distance=1.1cm and 3.2cm,on grid,auto]
		\tikzset{every state/.style={minimum size=9mm, inner sep=0}};
		\node[state] (q_0)  [initial, fill=orange!20] {$q_0$};
		\node[state] (s_0)  [above right=of q_0, text=red] {\small $\{1\}$};
		\node[state] (s_1)  [right=of q_0, text=red] {\small $\{0\}$};
		\node[state] (s_2)  [below=of s_1, text=red] {\small $\{2\}$};
		\node[state] (p_0)  [right=of s_0, text=red, fill=orange!20, label={right:{\footnotesize\color{blue!80} behave like} $q_0$}] {\small $\{0\}$};
		\node[state] (p_1)  [right=of s_1, text=red, fill=orange!20, label={right:{\footnotesize\color{blue!80} behave like} $q_0$}] {\small $\{1\}$};
		\node[state] (p_2)  [right=of s_2, text=red, fill=orange!20, label={right:{\footnotesize\color{blue!80} behave like} $q_0$}] {\small $\{2\}$};

		\path[->]
		(q_0) edge [sloped, pos=0.7, out=45, in=180] node {$1\in\sigma$} (s_0)
		(q_0) edge [sloped, pos=0.5] node {$0\in\sigma, 1\notin\sigma$} (s_1)
		(q_0) edge [sloped, pos=0.7, out=-45, in=180] node {$0,1\notin\sigma$} (s_2)
		(s_0) edge [] node {$\Sigma$} (p_0)
		(s_1) edge [] node {$\Sigma$} (p_1)
		(s_2) edge [] node {$\Sigma$} (p_2);
	\end{tikzpicture}
	\caption{Transducer $\cT^\pi$ for the $\cT$ in~\cref{example:rsym} and $\pi=(0\ 1)$.}
	\label{fig:example_transducer_pi}
\end{figure}

Consider a transducer $\cT$, we obtain from $\cT$ a new transducer $\cT^\pi$ by applying the permutation $\pi$ to the actions and labels. Formally, $\cT^{\pi}=\tup{2^\cP,2^\cP,Q,q_0,\delta^{\pi},\lab^{\pi}}$ where $\delta^{\pi}(q,\sigma)=\delta(q,\pi^{-1}(\sigma))$ and $\lab^{\pi}(q)=\pi(\lab(q))$. It is easy to verify that for every $x\in (2^{\cP})^*$ we have $\cT^\pi(x) = \pi(\cT(\pi^{-1}(x)))$.
\Cref{fig:example_transducer_pi} shows the transducer $\cT^\pi$ that corresponds to $\cT$ of~\cref{example:rsym} for $\pi=(0\ 1)$.

Once we have $\cT^\pi$, round symmetry can be expressed as round simulation, so we can use the tools developed in~\cref{sec:deciding_fixed_round_sim,sec:deciding_existential_round_sim} to solve the problems at hand.
\begin{lem}
	\label{lem:symmetry_to_simulation}
	For a permutation $\pi$ and $k>0$, $\cT$ is $k$-round symmetric \WRT $\pi$ iff $\cT^{\pi}\prec_k \cT$.
\end{lem}
\proof
	By definition, we have that $\cT^{\pi}\prec_{k} \cT$ iff for every $x\in (2^{\cP})^*$ there exists $x'\req x$ such that $\cT^{\pi}(x)\req \cT(x')$. We show that this is equivalent to the definition of round symmetry.

	For the first direction, assume $\cT$ is $k$-round symmetric \WRT $\pi$, and let $x\in (2^{\cP})^*$. Applying the definition of $k$-round symmetry to $y=\pi^{-1}(x)$, there exists $x'\req \pi(y)$ such that $\pi(\cT(y))\req \cT(x')$. Since $\pi(y)=x$ we get that $x'\req x$ and $\pi(\cT(\pi^{-1}(x)))\req \cT(x')$. By the above, $\cT^\pi(x) = \pi(\cT(\pi^{-1}(x)))$, so we have $\cT^{\pi}(x)\req \cT(x')$.

	For the second direction, assume $\cT^\pi \prec_k \cT$, and let $x\in (2^{\cP})^*$. Applying the definition of round simulation to $z=\pi(x)$, there exists $x'\req z$ such that $\cT^{\pi}(z)\req \cT(x')$. Thus, $\pi(\cT(\pi^{-1}(z)))\req \cT(x')$, but $\pi^{-1}(z)=x$, so we get $\pi(\cT(x))\req \cT(x')$, and we are done.
\qed

\paragraph*{Closure under composition.}
\Cref{lem:symmetry_to_simulation} enables us to naively solve fixed round symmetry by checking against all permutations. 
We show, however, that the definition above is closed under composition of permutations, allowing us to establish round symmetry by checking only two permutations, forming a generating set of $\cS_n$.
\begin{lem}
	\label{lem:closure_composition}
	Consider two permutations $\pi,\chi$. If $\cT^\pi\prec_k \cT$ and $\cT^\chi \prec_k \cT$ then $\cT^{\pi\circ \chi} \prec_k \cT$.
\end{lem}
\proof
	Using the first definition of round symmetry, let $x\in (2^{\cP})^*$, then there exists $x'\req[k] \pi(x)$ such that $\cT(x')\req[k] \pi(\cT(x))$. Moreover, there exists $x''\req[k] \chi(x') \req[k] \chi(\pi(x))$ such that $\cT(x'')\req[k] \chi(\cT(x'))\req[k] \chi(\pi(\cT(x)))$, and we are done.
\qed

Recall that the group of all permutations of $\cP=\{1,\ldots,m\}$ is generated by two permutations: the transposition $(1\ 2)$ and the cycle $(1\ 2\ \cdots\ m)$~\cite{Cameron1999}. By~\cref{lem:closure_composition} it is sufficient to check symmetry for these two generators in order to obtain symmetry for every permutation. Note that for the existential variant of the problem, even if every permutation requires a different $k$, by taking the product of the different values we conclude that there is a uniform $k$ for all permutations.
We thus have the following.
\begin{thm}
	\label{thm:symmetry_decidable}
	Both fixed and existential round symmetry are decidable. Moreover, fixed round symmetry is in \PSPACE.
\end{thm}

Finally, the reader may notice that our definition of round symmetry \WRT $\pi$ is not symmetric, as was the case with round simulation compared to round equivalence. However, when we consider round symmetry \WRT to all permutations, the definition becomes inherently symmetric, as a consequence of~\cref{lem:closure_composition}.
\begin{lem}
	\label{lem:round_symmetry_commutative}
	In the notations above, if $\cT^{\pi}\prec_k \cT$ then $\cT\prec_k \cT^\pi$.
\end{lem}
\proof
	Recall that for every permutation $\pi$ we have $\pi^{m!}=\mathtt{id}$, where $\mathtt{id}$ is the identity permutation. In particular, $\pi^{m!-1}=\pi^{-1}$.

	By~\cref{lem:closure_composition}, we now have that if $\cT^{\pi}\prec_k \cT$, then $\cT^{\pi^{m!-1}}\prec_k \cT$, so $\cT^{\pi^{-1}}\prec_k \cT$. Applying $\pi$ to both sides gives us $\cT\prec_k \cT^{\pi}$.
\qed

Thus, for symmetry, the notions of round simulation and round equivalence coincide.

\section{The Simulation Mapping}
\label{sec:equiv_mapping}
The definition of round simulation in~\cref{sec:round-equivalence} has an existential flavour: given input $x$ we consider the existence of a word $x'$ that satisfies the requirement of round simulation. In some cases it may be desirable to compute an $x'$ that ``witnesses'' the simulation of $x$. 

For example, recall the monitor of \cref{example:MC_rounds} modelled by a transducer $\cT_1$. Recall that we presented a simpler transducer $\cT_2$ that round-simulates $\cT_1$. This allowed us then to verify e.g., the property ``if there is no \texttt{error}, then Process $3$ works at least once every 20 steps'' against the much smaller $\cT_2$. 
When a designer wishes to gain understanding as to why the verification on $\cT_2$ is sound, they may want to see how input sequences/output sequences for $\cT_1$ are translated to $\cT_2$. In this example, the transformation is simple, and consists of ordering the process by their id.



Clearly one can compute $x'$ from $x$ by simply trying all permutations of $x$ and finding a successful one. This, however, is expensive, and raises the question of whether we can output $x'$ using a finite-state transducer. 
Unfortunately, we show in the following that computing $x'$ cannot be done locally, in the sense that arbitrary lookahead is needed.

Consider two transducers $\cT_1$ and $\cT_2$ such that $\cT_1 \prec_{k} \cT_2$, and an input word $x\in \left(\sI^k\right)^*$. This means, by definition, that there is a way to permute the rounds in $x$ to obtain a word $x'$ such that $\cT_2(x')$ is a permutation of $\cT_1(x)$.
A \emph{simulation~mapping\footnote{We omit $\Lambda$ for brevity. However, it can easily be incorporated.} between $\cT_1$ and $\cT_2$} is a function $\psi_{\cT_1,\cT_2}: \Sigma^* \to \Sigma^*$ such that for every $x\in \Sigma^{kR}$ we have that $x'=\psi_{\cT_1,\cT_2}(x)$ satisfies $x'\req[k]x$ and $\cT_1(x)\req[k]\cT_2(x')$ (we omit the subscripts when the transducers are clear from context). 

We start by showing that the simulation mapping is not a morphism, in the sense that it cannot act on each round separately.

\begin{exa}
\label{example:mapping}
Consider the transducers $\cT_1$ and $\cT_2$ depicted in~\cref{fig:example_mapping}, with input and output alphabets $\sI=\{a,b\}$ and $\sO=\{0,1\}$ and round length 2. $\cT_1$ expects to see either $ab$ or $ba$ in every 2-round, outputting $00$ in both cases, and otherwise outputs $01$ in that round. $\cT_2$ expects the first round to be $ab$ and the second to be $ba$, otherwise outputs $01$ in the round not meeting expectations; and beginning from the third round, it behaves like $\cT_1$. We have that $\cT_1\prec_{2}\cT_2$ by a permutation that corrects the order of the letters in the first two rounds of the input. Moreover, we have $\psi(ab)= \psi(ba)=ab$ whereas $\psi(abba)=abba \neq \psi(ab) \cdot \psi(ba)$.

\begin{figure}[ht]
	\centering
	\begin{tikzpicture}[shorten >=1pt,node distance=1.5cm and 1.5cm,on grid,auto]
	    \tikzset{every state/.style={minimum size=7mm}};
	    \node[state] (q'_i) [initial above, text=red, fill=blue!10] {$0$};
        \node[state] (q'_ab) [right=of q'_i, text=red, fill=blue!10] {$0$};
        \node[state] (q'_ba) [below=of q'_i, text=red, fill=blue!10] {$0$};
        \node[state] (q'_f) [below=of q'_ab, text=red, fill=blue!10] {$1$};
        \path[->] 
        (q'_i)  edge [bend left=10] node {$a$} (q'_ab)
        (q'_i)  edge [bend left=10] node {$b$} (q'_ba)
        (q'_ab) edge [bend left=10] node {$a$} (q'_f)
        (q'_ab) edge [bend left=10] node {$b$} (q'_i)
        (q'_ba) edge [bend left=10] node {$a$} (q'_i)
        (q'_ba) edge [bend left=10] node {$b$} (q'_f)
        (q'_f)  edge [bend left=10] node {$a$} (q'_ab)
        (q'_f)  edge [bend left=10] node {$b$} (q'_ba);
	    
        \node[state] (q_0) [initial above, right=of q'_ab, text=red] {};
        \node[state] (q1_a) [right=of q_0, text=red, fill=red!10] {$0$};
        \node[state] (q1_b) [below right=of q_0, text=red, fill=red!10] {$0$};
        \node[state] (q1_ab) [right=of q1_a, text=red, fill=red!10] {$0$};
        \node[state] (q1_ba) [right=of q1_b, text=red, fill=red!10] {$1$};
        \node[state] (q2_b) [right=of q1_ab, text=red, fill=green!10] {$0$};
        \node[state] (q2_a) [right=of q1_ba, text=red, fill=green!10] {$0$};
        \node[state] (q2_ba) [right=of q2_b, text=red, fill=green!10] {$0$};
        \node[state] (q2_ab) [right=of q2_a, text=red, fill=green!10] {$1$};
        \node[state] (q_i) [right=of q2_ba, text=red, fill=blue!10] {$0$};
        \node[state] (q_ab) [right=of q_i, text=red, fill=blue!10] {$0$};
        \node[state] (q_ba) [right=of q2_ab, text=red, fill=blue!10] {$0$};
        \node[state] (q_f) [right=of q_ba, text=red, fill=blue!10] {$1$};
        \path[->] 
        (q_0)   edge [] node {$a$} (q1_a)
        (q_0)   edge [sloped, pos=0.2] node {$b$} (q1_b)
        (q1_a)  edge [sloped, pos=0.2] node {$a$} (q1_ba)
        (q1_a)  edge [] node {$b$} (q1_ab)
        (q1_b)  edge [] node {$a,b$} (q1_ba)
        (q1_ab) edge [sloped, pos=0.2] node {$a$} (q2_a)
        (q1_ab) edge [] node {$b$} (q2_b)
        (q1_ba) edge [] node {$a$} (q2_a)
        (q1_ba) edge [sloped, pos=0.2] node {$b$} (q2_b)
        (q2_a)  edge [] node {$a,b$} (q2_ab)
        (q2_b)  edge [] node {$a$} (q2_ba)
        (q2_b)  edge [sloped, pos=0.2] node {$b$} (q2_ab)
        (q2_ab) edge [out=-30, in=-50, sloped, looseness=2, below] node {$a$} (q_ab)
        (q2_ab) edge [] node {$b$} (q_ba)
        (q2_ba) edge [out=30,in=140, sloped] node {$a$} (q_ab)
        (q2_ba) edge [sloped, pos=0.2] node {$b$} (q_ba)
        (q_i)   edge [bend left=10] node {$a$} (q_ab)
        (q_i)   edge [bend left=10] node {$b$} (q_ba)
        (q_ab)  edge [bend left=10] node {$a$} (q_f)
        (q_ab)  edge [bend left=10] node {$b$} (q_i)
        (q_ba)  edge [bend left=10] node {$a$} (q_i)
        (q_ba)  edge [bend left=10] node {$b$} (q_f)
        (q_f)   edge [bend left=10] node {$a$} (q_ab)
        (q_f)   edge [bend left=10] node {$b$} (q_ba);
    \end{tikzpicture}
	\caption{The transducers $\cT_1$ (left) and $\cT_2$ (right) in~\cref{example:mapping}. The states of $\cT_2$ in red, green and blue manage the first, second and later rounds, respectively.}
	\label{fig:example_mapping}
\end{figure}
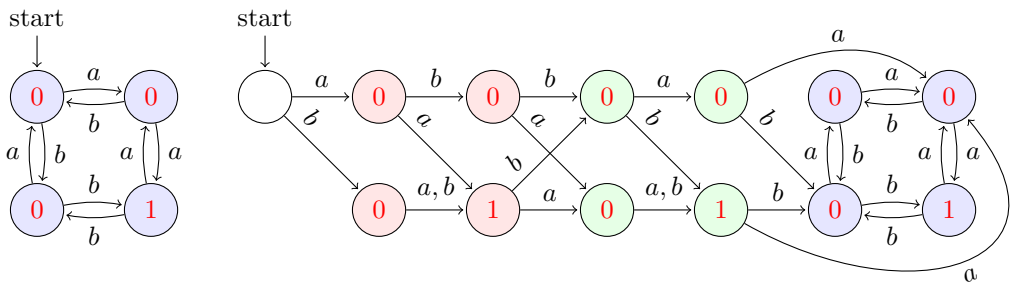
\end{exa}

Next, we show that in fact the simulation mapping cannot be described by any fixed lookahead machine.

\begin{exa}
\label{example:lookahead}
Set $\fL=L[ab\cdot (cc)^*\cdot (ab+ba)]$ and $k=2$, and let $\cT_1$ and $\cT_2$ be the transducers in~\cref{fig:example_lookahead}, satisfying $\cT_1\prec_{k,\fL} \cT_2$. Denote the simulation mapping by $\psi^*: (\sI^k)^* \rightarrow (\sI^k)^*$.

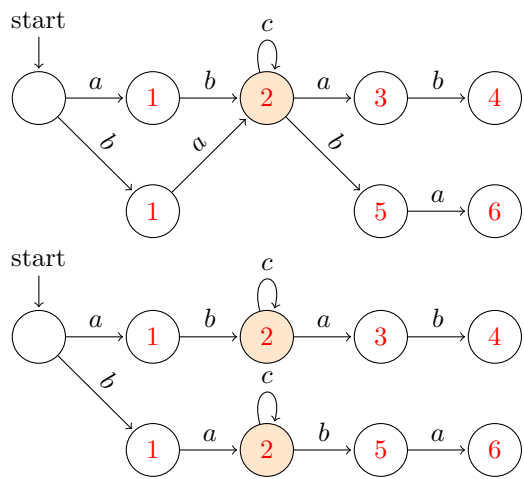
\begin{figure}[ht]
	\centering
	\begin{tikzpicture}[shorten >=1pt,node distance=1.5cm and 1.5cm,on grid,auto, scale=0.9]
	    \tikzset{every state/.style={minimum size=7mm,transform shape}};
	    \node[state] (q_0)  [initial above, text=red] {};
        \node[state] (q_a) [right=of q_0, text=red] {$1$};
        \node[state] (q_b) [below=of q_a, text=red] {$1$};
        \node[state] (q'_0) [right=of q_a, text=red, fill=orange!20] {$2$};
        \node[state] (q'_a) [right=of q'_0, text=red] {$3$};
        \node[state] (q'_b) [below=of q'_a, text=red] {$5$};
        \node[state] (q'_ab) [right=of q'_a, text=red] {$4$};
        \node[state] (q'_ba) [right=of q'_b, text=red] {$6$};
        \path[->] 
        (q_0)  edge [] node {$a$} (q_a)
        (q_0)  edge [sloped] node {$b$} (q_b)
        (q_a) edge [] node {$b$} (q'_0)
        (q_b) edge [sloped] node {$a$} (q'_0)
        (q'_0) edge [loop above] node {$c$} (q'_0)
        (q'_0) edge [] node {$a$} (q'_a)
        (q'_0) edge [sloped] node {$b$} (q'_b)
        (q'_a) edge [] node {$b$} (q'_ab)
        (q'_b) edge [] node {$a$} (q'_ba);
    \end{tikzpicture}%
    \quad 
	\begin{tikzpicture}[shorten >=1pt,node distance=1.5cm and 1.5cm,on grid,auto,scale=0.9]
	    \tikzset{every state/.style={minimum size=7mm,transform shape}};
	    \node[state] (q_0)  [initial above, text=red] {};
        \node[state] (q_a) [right=of q_0, text=red] {$1$};
        \node[state] (q_b) [below=of q_a, text=red] {$1$};
        \node[state] (q'_0) [right=of q_a, text=red, fill=orange!20] {$2$};
        \node[state] (q'_1) [right=of q_b, text=red, fill=orange!20] {$2$};
        \node[state] (q'_a) [right=of q'_0, text=red] {$3$};
        \node[state] (q'_b) [below=of q'_a, text=red] {$5$};
        \node[state] (q'_ab) [right=of q'_a, text=red] {$4$};
        \node[state] (q'_ba) [right=of q'_b, text=red] {$6$};
        \path[->] 
        (q_0)  edge [] node {$a$} (q_a)
        (q_0)  edge [sloped] node {$b$} (q_b)
        (q_a) edge [] node {$b$} (q'_0)
        (q_b) edge [] node {$a$} (q'_1)
        (q'_0) edge [loop above] node {$c$} (q'_0)
        (q'_1) edge [loop above] node {$c$} (q'_1)
        (q'_0) edge [] node {$a$} (q'_a)
        (q'_1) edge [] node {$b$} (q'_b)
        (q'_a) edge [] node {$b$} (q'_ab)
        (q'_b) edge [] node {$a$} (q'_ba);
    \end{tikzpicture}
	\caption{The transducers $\cT_1$ (left) and $\cT_2$ (right) in~\cref{example:lookahead}.}
	\label{fig:example_lookahead}
\end{figure}

We claim that for any $r$, there is no lookahead machine that defines a function $\psi_r: (\sI^{rk})^* \rightarrow (\sI^{rk})^*$ such that $\psi^*(x)=\psi(x)$ for all input words $x$.

Indeed, let $r\in\bbN$, and assume by way of contradiction that such $\psi_r$ exists. Now consider the input word $x=ab\cdot c^{rk-2}$. $\psi_r(x)$ must start with either $ab$ or $ba$. Without loss of generality, assume the former, and consider the input word $x':=x\cdot ba\cdot c^{rk-2}$. Since $\psi_r$ works on $r$ rounds each time, the first $r$ rounds are fixed when it reads the $(r+1)$-th round. Moreover, since $\psi_r(x')$ must induce a valid path in $\cT_2$, the only option for the $(r+1)$-th round of $\psi_r(x')$ is $ab$. Hence, the output of $\cT_1$ on $x'$ is different from the output of $\cT_2$ on $\psi(x')$, and we have a contradiction.
\end{exa}
\Cref{example:lookahead} essentially shows that it is generally impossible to determine the output of the first round without knowing the entire input. In \cref{sec:conclusion} we discuss possible models that may be able to capture it, and are weaker than general Turing machines.

\section{Additional Notions of Symmetry and Simulation}
\label{sec:other_notions}
Recall that under our definition from \cref{sec:preliminaries}, we have that $x\req[k]y$ if every $k$-round of $x$ can be permuted to a $k$-round of $y$. This permutation, however, can vary between rounds. In some settings, we would want the rounds to be transformed uniformly, with the same permutation. To this end, we introduce below the notion of \emph{uniform round simulation}. In addition, if the underlying alphabet consists of set of signals, as in the setting of \cref{sec:symmetry_to_simulation}, we can also consider simulation where one is allowed to permute the index of each signal, instead of entire letters. To capture this notion, we introduce \emph{signal-wise simulation}. Finally, recall that simulation is defined by permutation of both the input and output letters. Given the new definitions, one can consider simulations where the inputs and outputs are not similarly permuted, e.g., the inputs can be permuted arbitrarily, but the outputs need to be permuted uniformly.
In the following, we discuss these notions and their interrelations.

For brevity, we omit $\fL$ from this discussion, as it is an orthogonal restriction and can be easily incorporated to the setting.

\subsection{Variations of Round Symmetry and Round Simulation}
\label{sec:variations_rs}
\newcommand{\RSTYPES}{\ensuremath{\mathbf{MOP}}}
\newcommand{\RSTUPS}{\ensuremath{\mathbf{MOP}^2_k}}
\newcommand{\rstype}[1]{\mathtt{#1}}
\newcommand{\reqtype}[2][]{\req[#1]^{\rstype{#2}}}
\newcommand{\prectype}[2]{\prec^{\rstype{#1}, \rstype{#2}}}
We start by formally defining new notions of simulation. For this section, we consider $2^\cP/2^\cP$ transducers\footnote{the choice of $2^\cP$ as both the input and output alphabet is arbitrary.} for $\cP=\{1,\ldots, n\}$.

Consider two words $x,y\in (2^\cP)^*$ of length $kR$. We say that $x,y$ are \emph{uniformly round equivalent} and denote by $x\reqtype[k]{u} y$ if $x\req[k]y$ and there exists a single permutation $\tau\in \cS_k$ which transforms the rounds of $x$ to those of $y$.
We say that $x,y$ are \emph{signal-wise round equivalent}, denoted $x\reqtype[k]{s} y$, if for each $k$-round, $x$ and $y$ have the same number of occurrences of each signal. More precisely, for each signal $p\in \cP$ and round $0\le i<R$, we have $|\{j: p\in x_{ik+j}, 1\le j\le k\}|=|\{j: p\in y_{ik+j}, 1\le j\le k\}|$. For clarity, we explicitly denote our original definition of round equivalence by  $x\reqtype[k]{\ell} y$, where $\ell$ stands for ``letter'' round equivalence.
We refer to the three types of round equivalence as \emph{modes}.

The new definitions lift to simulation of transducers, by specifying which type of round equivalence is used on the inputs and outputs. We thus obtain 9 definitions of simulation, as follows. Consider transducers $\cT_1,\cT_2$, and let $\mu,\mu'\in \{s,\ell,u\}$ be modes of round equivalence. We write $\cT_1 \prec_{k}^{\mu,\mu'} \cT_2$ if for every input $x$ there exists $x'\reqtype[k]{\mu}x$ such that $\cT_1(x)\reqtype[k]{\mu'}\cT_2(x')$. This definition is in turn lifted to symmetry, as per \cref{sec:symmetry_to_simulation}, by replacing $\cT_2$ with $\cT^\pi$ for a permutation $\pi$ of the signals.

\begin{exa}[Round Robin is uniform symmetric]
Consider the RR scheduler for $n$ processes, shown to be $n$-round symmetric in~\cref{example:rr_rsym}. Recall that in the proof of its symmetry when the permutation $\pi$ was applied to the signals, we had to change the order of handling the requests such that it matched the new order of received requests: given input $x$, for the $i$-th round $b_1 b_2\cdots b_n$ of $\pi(x)$ (the input under permutation $\pi$) we set the corresponding round in $x'$ to $b_{\pi^{-1}(1)} b_{\pi^{-1}(2)} \cdots b_{\pi^{-1}(n)}$. Since the same permutation $\pi$ was applied for all rounds of the input $x$, the permutation by which the rounds of $x'$ were obtained was identical for all rounds. It follows that RR exhibits uniform round symmetry, i.e., $\cT \prec_{k}^{u,u} \cT^\pi$.
\end{exa}

\begin{figure}[ht]
	\centering
	\begin{tikzpicture}[shorten >=1pt,node distance=1.25cm and 1.25cm,on grid,auto]
		\node (top)    [] {$\tup{s,s,k}$}; 
		\node (topleft)   [below left=of top]   {$\tup{s,\ell,k}$};
		\node (topright)  [below right=of top]  {$\tup{\ell,s,k}$};
		\node (middle) [below right=of topleft] {$\tup{\ell,\ell,k}$};
		\node (botleft)   [below left=of middle]   {$\tup{\ell,u,k}$};
		\node (botright)  [below right=of middle]  {$\tup{u,\ell,k}$};
		\node (bottom) [below right=of botleft] {$\tup{u,u,k}$};
		\node (left)   [below left=of topleft]   {$\tup{s,u,k}$};
		\node (right)  [below right=of topright]  {$\tup{u,s,k}$};
		
 		\path[->] 
		(topleft)       edge node [] {} (top)
		(topright)       edge node [] {} (top)
		(middle)   edge node [] {} (topleft)
		(middle)  edge node [] {} (topright)
		(botleft)    edge node [] {} (middle)
		(botright)    edge node [] {} (middle)
		(bottom)   edge node [] {} (botleft)
		(bottom)  edge node [] {} (botright)
		(left)   edge node [] {} (topleft)
		(botleft)      edge node [] {} (left)
		(right)  edge node [] {} (topright)
		(botright)     edge node [] {} (right);
	\end{tikzpicture}
	\caption{A Hasse diagram for the partial order on the strictness of the definitions, where $\alpha\to \beta$ means  $\alpha$ implies $\beta$.}
	\label{fig:psu-diagram}
\end{figure}
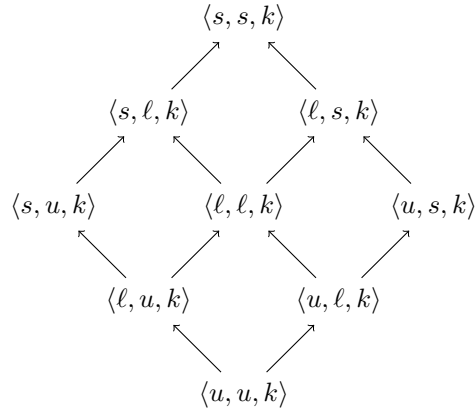

The modes of equivalence can be compared by their strictness, with uniform equivalence implying letter-wise, which in turn implies signal-wise. This can be lifted to round equivalence, yielding a partial order on the strictness of the various definitions, as depicted in \cref{fig:psu-diagram}.

In the remainder of the section, we give some examples separating some of the definitions, thus showing the order in \cref{fig:psu-diagram} is strict. Similar examples can be constructed for separating the rest of the definitions.

\begin{exa}[Symbol-wise symmetry does not imply letter-wise symmetry]
\label{example:parikh_not_symbolwise}
We warm up by showing that $\tup{\ell,\ell,k}$ is more strict than $\tup{s,s,k}$ (we will later reuse this example to establish finer strictness results).
Set $\pi=(0\ 1)$ and let $k\in \bbN$ and $m\geq 3$. We construct a transducer that is symbol-wise $k$-round symmetric, but not letter-wise $k'$-round symmetric for any $k'$.

Consider the $2^\cP/2^\cP$ transducer $\cT=\tup{2^\cP,2^\cP,S,s_0,\delta,\lab}$ depicted in~\cref{fig:example_parikh_not_symbolwise}, where $\cP=[m]=\{0,\cdots, m-1\}$.

\begin{figure}[ht]
	\centering
	\begin{tikzpicture}[shorten >=1pt,node distance=1.5cm and 1.8cm,on grid,auto]
	    \tikzset{every state/.style={minimum size=7mm}};
	    \node[state] (q_0)  [initial, fill=orange!20] {$q_0$};
        \node[state] (s_1) [above right=of q_0, text=red] {$\emptyset$};
        \node[state] (s_2) [right=of s_1, text=red] {$\emptyset$};
        \node[draw=none] (s_ellipsis) [right=of s_2] {$\cdots$};
        \node[state] (s_k-1) [right=of s_ellipsis, text=red] {$\emptyset$};
        \node[state] (s_k) [right=of s_k-1, text=red, fill=orange!20, label={right:{\footnotesize\color{blue!80} behave like} $q_0$}, inner sep=0] {\small $\{0\}$};
        \node[state] (t_1) [right=of q_0, text=red] {$\emptyset$};
        \node[state] (t_2) [right=of t_1, text=red] {$\emptyset$};
        \node[draw=none] (t_ellipsis) [right=of t_2] {$\cdots$};
        \node[state] (t_k-1) [right=of t_ellipsis, text=red] {$\emptyset$};
        \node[state] (t_k) [right=of t_k-1, text=red, fill=orange!20, label={right:{\footnotesize\color{blue!80} behave like} $q_0$}, inner sep=0] {\small $\{1\}$};
        \node[state] (p_1) [below right=of q_0, text=red, label={[xshift=0.4cm, yshift=-1cm]\footnotesize\color{black!50} 1}] {$\emptyset$};
        \node[state] (p_2) [right=of p_1, text=red, label={[xshift=0.4cm, yshift=-1cm]\footnotesize\color{black!50} 2}] {$\emptyset$};
        \node[draw=none] (p_ellipsis) [right=of p_2] {$\cdots$};
        \node[state] (p_k-1) [right=of p_ellipsis, text=red, label={[xshift=0.7cm, yshift=-1cm]\footnotesize\color{black!50} $k-1$}] {$\emptyset$};
        \node[state] (p_k) [right=of p_k-1, text=red, fill=orange!20, label={right:{\footnotesize\color{blue!80} behave like} $q_0$}, label={[xshift=0.4cm, yshift=-1cm]\footnotesize\color{black!50} $k$}] {$\emptyset$};
        
        \path[->] 
        (q_0) edge [sloped, pos=0.3, bend left] node {$\{0\}$} (s_1)
        (q_0) edge [] node {$\{1,2\}$} (t_1)
        (q_0) edge [sloped, pos=0.3, bend right] node {else} (p_1)
        (s_1) edge [] node {$\sI$} (s_2)
        (s_2) edge [] node {$\sI$} (s_ellipsis)
        (s_ellipsis) edge [] node {$\sI$} (s_k-1)
        (s_k-1) edge [] node {$\{1,2\}$} (s_k)
        (s_k-1) edge [sloped, pos=0.15] node {else} (p_k)
        (t_1) edge [] node {$\sI$} (t_2)
        (t_2) edge [] node {$\sI$} (t_ellipsis)
        (t_ellipsis) edge [] node {$\sI$} (t_k-1)
        (t_k-1) edge [] node {\hspace*{8pt}$\{0\}$} (t_k)
        (t_k-1) edge [sloped, pos=0.3] node {else} (p_k)
        (p_1) edge [] node {$\sI$} (p_2)
        (p_2) edge [] node {$\sI$} (p_ellipsis)
        (p_ellipsis) edge [] node {$\sI$} (p_k-1)
        (p_k-1) edge [] node {$\sI$} (p_k);
    \end{tikzpicture}
	\caption{$\cT$ exhibits symbol, but not letter-wise, round symmetry (see~\cref{example:parikh_not_symbolwise}).}
	\label{fig:example_parikh_not_symbolwise}
\end{figure}
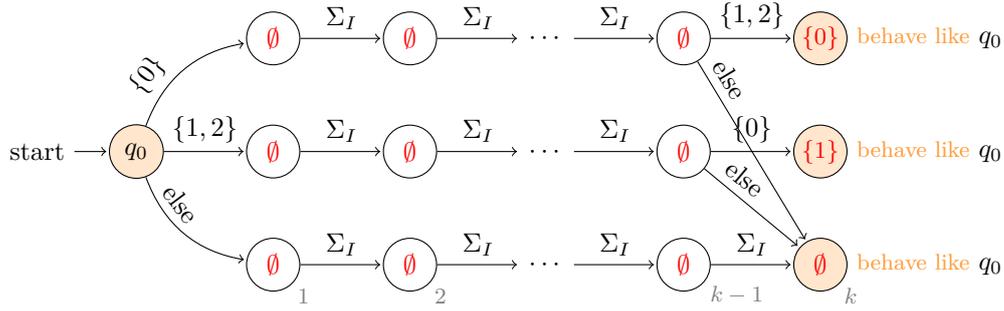

Observe that every round starts at $q_0$. There are three possible forms for the output of each round depending on the input, as summarized in~\cref{tab:example_parikh_not_symbolwise}.

\begin{table}[!htb]
    \centering
    \caption{The inputs and their corresponding outputs in $\cT$ of~\cref{example:parikh_not_symbolwise}.}
    \vspace{2mm}
    \def\arraystretch{1.3}
    \begin{tabular}{c|c}
        Input & Output \\
        \hline \hline
        $\{0\}\sigma_2\cdots \sigma_{k-1}\{1,2\}$ & $\emptyset^{k-1} \{0\}$ \\
        \hline
        $\{1,2\}\sigma_2\cdots \sigma_{k-1}\{0\}$ & $\emptyset^{k-1} \{1\}$ \\
        \hline
        else & $\emptyset^k$ \\
    \end{tabular}
    \label{tab:example_parikh_not_symbolwise}
\end{table}

We first show that $\cT$ is symbol-wise round symmetric. Let $x$ be an input word.
Similarly to~\cref{sec:application}, $\pi(x)$ is the word obtained from $x$ by permuting every signal according to $\pi$.
If $x$ is of one of the first two forms in~\cref{tab:example_parikh_not_symbolwise}, then by moving the signal $2\in \cP$ (fixed in $\pi$) between the first and last letters, we get $x'\reqtype[]{s} \pi(x)$ such that $T(x')\reqtype[]{s} \pi(T(x))$, as desired. Now assume $x$ is of some other form, having the output $\emptyset^k$. If $2\in \cP$ appears in both the first and last letters, or it appears in neither, then set $x'=\pi(x)$; otherwise, move the signal 2 to the other letter, and the output will remain $\emptyset^k$. Thus, $\cT$ is symbol-wise round symmetric.

On the other hand, $\cT$ is not letter-wise $k'$-round symmetric for any $k'>0$. To see this, take the input $x=\{0\}^{k-1}\cdot\{1,2\}\cdot\emptyset^{k'k-k}$. We have $|x|=k'k$ which is divisible by $k'$, $\cT(x)=\emptyset^{k-1}\cdot\{0\}\cdot\emptyset^{k'k-k}$. It holds that $\pi(x)=\{1\}^{k-1}\cdot\{0,2\}\cdot\emptyset^{k'k-k}$, which contains neither the letter $\{0\}$ nor $\{1,2\}$. Thus, regardless of how we permute $\pi(x)$ to obtain $x'$, the output of any $x'\reqtype[]{\ell}\pi(x)$ is always $\emptyset^{k'k}$, which is not a permutation of $\cT(x)$.
\end{exa}

\begin{exa}[Showing $\tup{\rstype{s}, \rstype{s}, k}\lneq\tup{\rstype{\ell}, \rstype{s}, k}$]
\label{example:gap1}
Let $\cT$ be the transducer from~\cref{example:parikh_not_symbolwise}, and consider the transducer $\cT^\pi$ obtained from $\cT$ by permuting both the input and the output by $\pi=(0\ 1)$ as in~\cref{sec:symmetry_to_simulation}. We have shown that $\cT$ is symbol-wise round symmetric. By a reasoning analogous to the transition from symmetry to simulation as per~\cref{sec:symmetry_to_simulation}, this gives $\cT\prectype{s}{s}_k \cT^\pi$. However, it does not hold that $\cT\prectype{\ell}{s}_k \cT^\pi$: for the input $x:=\{0\}\sigma_2\cdots \sigma_{k-1}\{1,2\}$ having output $y:=\emptyset^{k-1}\{0\}$ (cf.~\cref{tab:example_parikh_not_symbolwise}), any permutation $x'\reqtype[k]{\ell} x$ will lead to an output of $\emptyset^k\not\reqtype[k]{s} y$. Thus $\cT\not\prectype{\ell}{s}_k \cT^\pi$ (and in particular, $\cT\not\prectype{\ell}{\ell}_k \cT^\pi$ so $\cT$ is not letter-wise symmetric). In the general sense, we conclude that $\cT_1\prectype{s}{s}_k \cT_2$ does not imply $\cT_1\prectype{\ell}{s}_k \cT_2$.
\end{exa}


\begin{exa}[Showing $\tup{\rstype{s}, \rstype{s}, k}\lneq\tup{\rstype{s}, \rstype{\ell}, k}$]
\label{example:gap2}
Consider the transducer $\cT$ in~\cref{fig:example_gap2}, whose round-by-round behaviour can once more be summarized in a table (see~\cref{tab:example_gap2}).
$\cT$ is symbol-wise round symmetric: for an input $x$, choose $x'=\pi(x)$. It is not difficult to show that $\cT(x')\reqtype[k]{s} \pi(\cT(x))$ by considering the possible forms of $x$ according to~\cref{tab:example_gap2}.
To see that $\cT\nprec^{s,\ell}_k \cT^\pi$, consider the word $x=\{0\} \emptyset \emptyset$. The output of $\cT$ on $x$ is $\{0\}\emptyset \{2\}$. Any round equivalent word $x'$ of $x$ either starts with $\{1\}$ or $\emptyset$, the respective outputs being either $\{1,2\}\emptyset\emptyset$ or $\emptyset^3$. In all cases, we have $T(x')\not\reqtype[k]{\ell} \cT^{\pi}(x)$.

\begin{figure}[ht]
	\centering
	\begin{tikzpicture}[shorten >=1pt,node distance=1.1cm and 3.1cm,on grid,auto]
	    \tikzset{every state/.style={minimum size=9mm, inner sep=0}};
	    \node[state] (q_0)  [initial, fill=orange!20] {$q_0$};
	    \node[state] (s_0)  [above right=of q_0, text=red] {\small $\{0\}$};
	    \node[state] (s_1)  [below right=of q_0, text=red] {\small $\{1\}$};
	    \node[state] (s_2)  [below=of s_1, text=red] {\footnotesize $\{0,2\}$};
	    \node[state] (s_3)  [below=of s_2, text=red] {\footnotesize $\{1,2\}$};
	    \node[state] (s_4)  [below=of s_3, text=red] {$\emptyset$};
	    \node[state] (t_0)  [right=of s_0, text=red] {$\emptyset$};
	    \node[state] (t_1)  [right=of s_1, text=red] {\small $\{2\}$};
	    \node[state] (t_2)  [right=of s_3, text=red] {$\emptyset$};
	    \node[state] (p_0)  [right=of t_0, text=red, fill=orange!20, label={right:{\footnotesize\color{blue!80} behave like} $q_0$}] {\small $\{2\}$};
	    \node[state] (p_1)  [right=of t_1, text=red, fill=orange!20, label={right:{\footnotesize\color{blue!80} behave like} $q_0$}] {\small $\emptyset$};
	    \node[state] (p_2)  [right=of t_2, text=red, fill=orange!20, label={right:{\footnotesize\color{blue!80} behave like} $q_0$}] {\small $\emptyset$};
        
        \path[->] 
        (q_0) edge [sloped, pos=0.7] node {$\{0\}$} (s_0)
        (q_0) edge [sloped, pos=0.7] node {$\{1,2\}$} (s_1)
        (q_0) edge [sloped, pos=0.7, bend right] node {$\{0,2\}$} (s_2)
        (q_0) edge [sloped, pos=0.7, bend right] node {$\{1\}$} (s_3)
        (q_0) edge [sloped, pos=0.7, bend right] node {else} (s_4)
        (s_0) edge [sloped, pos=0.3] node {$2\notin \sigma$} (t_0)
        (s_0) edge [sloped, pos=0.3] node {$2\in \sigma$} (t_1)
        (s_1) edge [sloped, pos=0.3] node {$2\notin \sigma$} (t_1)
        (s_1) edge [sloped, pos=0.2] node {$2\in \sigma$} (t_0)
        (s_2) edge [sloped, pos=0.3] node {$\Sigma$} (t_2)
        (s_3) edge [sloped, pos=0.3] node {$\Sigma$} (t_2)
        (s_4) edge [sloped, pos=0.3] node {$\Sigma$} (t_2)
        (t_0) edge [] node {$\Sigma$} (p_0)
        (t_1) edge [] node {$\Sigma$} (p_1)
        (t_2) edge [] node {$\Sigma$} (p_2);
    \end{tikzpicture}
	\caption{The transducer $\cT$ for~\cref{example:gap2}. The transitions $i\in\sigma$ and $i\notin\sigma$ mean all letters from $\sI$ that, respectively, contain or do not contain $i$.}
	\label{fig:example_gap2}
\end{figure}
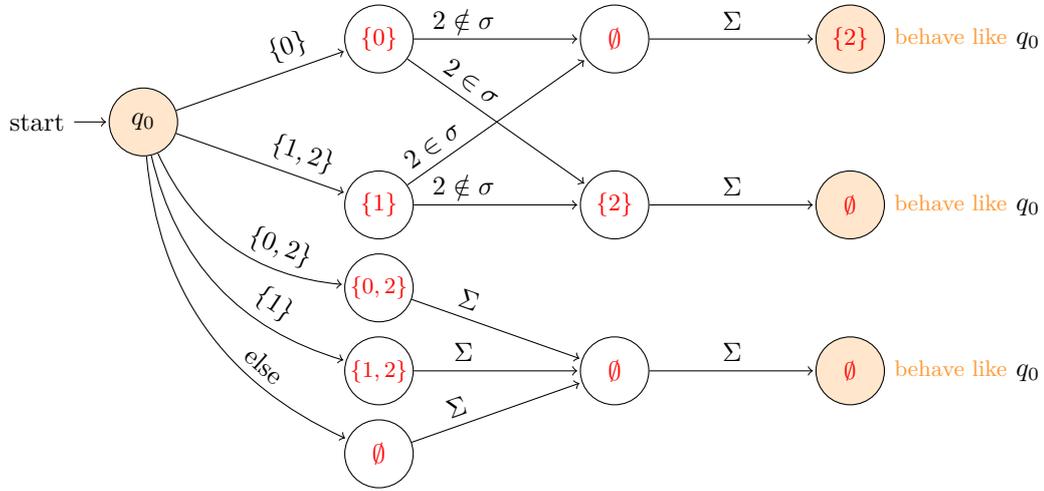

\begin{table}[!htb]
    \centering
    \caption{The inputs and their corresponding outputs in $\cT$ of~\cref{example:gap2}.}
    \vspace{2mm}
    \def\arraystretch{1.3}
    \begin{tabular}{c|c}
        Input & Output \\
        \hline \hline
        $\{0\}(2\notin\sigma)\,\sigma$ & $\{0\}\emptyset\{2\}$ \\
        \hline
        $\{0\}(2\in\sigma)\,\sigma$ & $\{0\}\{2\}\emptyset$ \\
        \hline
        $\{1,2\}(2\in\sigma)\,\sigma$ & $\{1\}\emptyset\{2\}$ \\
        \hline
        $\{1,2\}(2\notin\sigma)\,\sigma$ & $\{1\}\{2\}\emptyset$ \\
        \hline
        $\{0,2\}\sigma\sigma$ & $\{0,2\}\emptyset\emptyset$ \\
        \hline
        $\{1\}\sigma\sigma$ & $\{1,2\}\emptyset\emptyset$ \\
        \hline
        else & $\emptyset\emptyset\emptyset$ \\
    \end{tabular}
    \label{tab:example_gap2}
\end{table}
\end{exa}

The transducers used in~\cref{example:gap1,example:gap2} have established two gaps from~\cref{fig:psu-diagram}. In fact, these same transducers can be used to establish the remaining two dual gaps as well, as follows. The transducer $\cT$ in~\cref{example:gap1} satisfies $\tup{\rstype{s},\rstype{\ell},k}$-round simulation with its corresponding $\cT^\pi$; indeed, observe that the output labels are either singleton sets or empty sets, so that a signal permutation of the output is equivalent to permuting the letters. The transducer $\cT$ in~\cref{example:gap2} satisfies $\tup{\rstype{\ell},\rstype{s},k}$-round simulation with its corresponding $\cT^\pi$, which is inferred from the choice of $x'=\pi(x)$, satisfying in particular $x'\reqtype[]{\ell} \pi(x)$. However, neither of the two satisfy $\tup{\rstype{\ell},\rstype{\ell},k}$-round simulation, since they are not symbol-wise round symmetric. This completes the proof of strictness of top diamond in~\cref{fig:psu-diagram}.
In~\cref{apx:variants} we provide constructions to complete some of the remaining strictness results.

Finally, \cref{apx:example_counterex} presents a pair of transducers $\cT_1$ and $\cT_2$ such that $\cT_1\prectype{\ell}{s}_2 \cT_2$ and $\cT_1\prectype{s}{\ell}_2 \cT_2$, but $\cT_1\not\prectype{\ell}{\ell}_2 \cT_2$. This proves that although $\tup{\rstype{\ell},\rstype{\ell},k}$-round simulation implies both $\tup{\rstype{s},\rstype{\ell},k}$ and $\tup{\rstype{\ell},\rstype{s},k}$-round simulation, the converse does not hold.

\subsection{Deciding Round Simulation}
\label{sec:decision_new_notions}
We briefly discuss the decidability of round simulation for the new notions. 
We start by considering $\tup{\rstype{s},\rstype{s},k}$-round simulation, where the following arguments also apply when replacing one of the $\rstype{s}$ with $\rstype{\ell}$. 
The main idea is to tweak the definitions of~\cref{sec:deciding_fixed_round_sim,sec:deciding_existential_round_sim}, and specifically the permutation-closure NFA, to look at permutations of the signals, not just the letters.
To this end, we simply modify the notion of \emph{Parikh image} over an alphabet $2^{\cP}$ to be with respect to $\cP$. That is, for $x\in (2^{\cP})^*$, let $\fP(x)\in \bbN^\cP$ be the vector counting the number of occurrences of each signal $p\in \cP$ in the letters of $x$.

Under this definition, the analysis of \cref{sec:deciding_fixed_round_sim,sec:deciding_existential_round_sim} follows without any changes. Indeed, the crucial property that is needed for these arguments is that the permutation-closure NFA is indeed closed under permutation, which clearly holds also for the new definition. In particular, the proof of~\cref{lem:round_equivalence_iff_perm_containment} hold, from which the rest of the analysis follows.
Thus, adding $\rstype{s}$ to the model retains the decidability and complexity of both fixed round simulation and existential round simulation.

In contrast, uniform round simulation is conceptually different: the constraint on the permutations of each round is now global for the word. That is, we need a single permutation to be used in all rounds. This means that the techniques of \cref{sec:deciding_fixed_round_sim,sec:deciding_existential_round_sim} no longer apply.
Moreover, uniform round simulation is \emph{not} invariant to (letter or signal) round permutations. Indeed, clearly there are words $x\reqtype[k]{\ell} x'$ and $y$ such that $x\reqtype[k]{u} y$ but $x'\not\reqtype[k]{u} y$. 

For fixed round simulation, enforcing the global condition is not too difficult, as we now show.
\begin{thm}
	\label{thm:fixed_uniform_re_PSPACE}
	Given transducers $\cT_1,\cT_2$ and $k>0$ in unary, the problem of deciding whether $\cT_1 \prec_{k,}^{u,u} \cT_2$ is in \PSPACE.
\end{thm}
\proof
    Recall that $\cT_1 \prec_{k}^{u,u} \cT_2$ iff for every $x$ there exist permutations $\pi,\tau$ such that  $\pi(x)=y$ (where $\pi(x)$ is the word obtained by applying $\pi$ to each $k$-round of $x$) and $\tau(\cT_1(x))=\cT_2(y)$.

    Let $\cD^k_1$ and $\cD^k_2$ be the trace DFAs of $\cT_1$ and $\cT_2$ as per~\cref{sec:deciding_fixed_round_sim}, where we modify them to read the alphabet $\sI^k\times \sO^)$ (in this setting $\sI=\sO=2^{\cP}$). 
    Next, for permutations $\pi,\tau$ as above, define $\cA^{\pi,\tau}_1$ to be the DFA obtained from $\cD^k_1$ by, intuitively, applying $\pi,\tau$ to $\sI^k\times\sO^k$. Formally, let $\delta:Q\times(\sI^k\times\sO^k)\to Q$ be the transition function of $\cD^k_1$, then the transition function of $\cA^{\pi,\tau}_1$ is given by $\mu(q,(\alpha,\beta))=\delta(q,(\pi(\alpha),\tau(\beta)))$.
    We now obtain an NFA $\cA$ by taking the union of $\cA^{\pi,\tau}_1$ over all permutations $\pi,\tau$. It is easy to see that $\cT_1 \prec_{k}^{u,u} \cT_2$ iff $L(\cA)\subseteq \cD^k_2$.

    Since the size of $\cA$ is single-exponential in that of $\cD^k_1$, but can be construction on-the-fly, the latter containment can be decided in \PSPACE.
\qed

\cref{thm:fixed_uniform_re_PSPACE} can be easily combined with the remaining notions to obtain the decidabilty of all nine definitions of fixed round simulation.

\begin{rem}
    \label{rmk:uniform_existential}
    Unfortunately, the construction in the proof of~\cref{thm:fixed_uniform_re_PSPACE} significantly modifies the state space of $\cD^k_1$. This is in contrast to the construction in~\cref{lem:round_equivalence_iff_perm_containment}, which only modifies the transition function.

    In particular, it is not clear if the construction can be symbolically defined via e.g., Presburger Arithmetic (or some other decidable logic) in order to extend decidability to the existential-bound setting. We therefore leave the latter as an open problem.
\end{rem}

\section{Conclusion and Open Questions}
\label{sec:conclusion}
\label{sec:discussion}
\label{sec:future}
 
In this work, we introduced round simulation and provided decision procedures and lower bounds (some with remaining gaps) for the related algorithmic problems.  Our framework can be viewed as a notion of ``approximate simulation'', by which we can significantly reduce the state space for verification, at the cost of invariance to permutations.

Round simulation, and in particular its application to round symmetry, is only an instantiation of a more general framework of symmetry, by which we measure the stability of transducers under local changes to the input. In particular, there is place for additional notions of symmetry and simulation to be studied, and the existing ones extended. Some such variants were presented and discussed in~\cref{sec:variations_rs}, but others, e.g., sliding-window symmetry, or the setting of infinite words may also be of interest in future works.

A few gaps have remained open in this work. Most notably are tightening the complexity gap of existential simulation~\cref{rmk:complexity}, and implementing the simulation mapping from~\cref{sec:equiv_mapping} using a simpler computational model than Turing machines. Some possible candidates for the latter are streaming-string transducers and bi-machines~\cite{Muscholl}.

\bibliography{main.bib}

\appendix
\section{PSPACE Hardness}
\label{chap:apx}

\begin{lem}
\label{lem:universalityofnfa}
Universality of \NFAs over alphabet $\Sigma=\{0,1\}$, where all states are accepting, and the degree of nondeterminism is at most $2$, is $\PSPACE$-complete.
\end{lem}
\proof
In~\cite{kao2009nfas}, it is shown that universality of \NFAs remains $\PSPACE$-complete even for \NFAs over alphabet $\Sigma=\{0,1\}$ and all states accepting. Thus, we only need to show that this remains the case under the restriction that $|\delta(q,\sigma)|\le 2$ for every state $q$ and letter $\sigma$.

To see this, we start by observing that universality remains $\PSPACE$-complete for \NFAs over alphabet $\{0,1,\$\}$ with nondeterminism degree at most 2. Indeed, given an \NFA over $\{0,1\}$ with maximal nondeterminism degree $d>2$, we can replace each transition of the form\footnote{We can assume all transitions have degree exactly $d$ by adding redundant transitions} $\delta(q,\sigma)=\{q_1,\ldots, q_d\}$ with a binary tree of depth $\lceil \log d \rceil$, reading $\$$ on all transitions, which starts at $q$ and ends in $q_1,\ldots,q_d$. Thus, we introduce at most $d$ states for every transition. By marking these states as accepting, this reduction maintains universality, and requires a polynomial blowup.

Next, we observe that the reductions in~\cite[Lemma 2]{kao2009nfas} first transform an \NFA over alphabet size $k$ to an \NFA over alphabet size $k+1$ with all states accepting and with identical nondeterminism degree (indeed, the only added transitions are in fact deterministic), and then transforms an \NFA with all states accepting and alphabet size $4$ to an \NFA with all states accepting and alphabet size $2$, with an equal nondeterminism degree (essentially by encoding each of the 4 letters as two letters in $\{0,1\}$).

Since we start this chain of reductions with an \NFA of nondeterminism degree at most 2, we maintain this property throughout the proof.
\qed

\subsection{Proof of Theorem~\ref{thm:equivalence_PSPACE-H}}
\label{apx:proof_equivalence_PSPACE-H}

We show a reduction from the universality problem for \NFAs over alphabet $\{0,1\}$ where all states are accepting and the degree of nondeterminism is at most 2, to round equivalence with $k=2$ and with $\fL$ given as a \DFA of constant size. The former is shown to be $\PSPACE$-hard in~\cref{lem:universalityofnfa}.

Consider an \NFA $\cN=\tup{Q,\{0,1\},\delta,q_0,Q}$ where $|\delta(q,\sigma)|\le 2$ for every $q\in Q$ and $\sigma\in \{0,1\}$.
We construct two transducers $\cT_1$ and $\cT_2$ over input and output alphabets $\sI=\{a,b,c,d\}$ and $\sO=\{\top,\bot\}$ and $\fL\subseteq \sI^*$, such that $L(\cN)=\{0,1\}^*$ iff $\cT_1\equiv_{2,\fL}\cT_2$. 

Set $\fL=(ab+cd)^*$ (described as a 4-state \DFA). Intuitively, our reduction encodes $\{0,1\}$ into $\{a,b,c,d\}^2$ by setting $0$ to correspond to $ab$ and to $ba$, and $1$ to $cd$ and to $dc$. Then, $\cT_1$ keeps outputting $\top$ for all inputs in $\fL$, thus mimicking ``accepting'' every word in $\{0,1\}^*$. We then construct $\cT_2$ so that every nondeterministic transition of $\cN$ on e.g., $0$ is replaced by two deterministic branches on $ab$ and on $ba$. Hence, when we are allowed to permute $ab$ and $ba$ by round equivalence, we capture the nondeterminism of $\cN$. 

\begin{figure}[ht]
\begin{minipage}{.31\linewidth}
	\begin{tikzpicture}[shorten >=1pt,node distance=1.5cm and 1.3cm,on grid,auto]
	    \tikzstyle{state}=[circle, inner sep=0mm, outer sep=0mm, minimum size=8mm, draw=black];
		\node[state] [initial text={\,}] (q_0) [red, draw=black, initial above] {$\top$}; 
		\node[state] (q_1) [red, draw=black, right=of q_0] {$\top$}; 
		\node[state] (q_2) [red, draw=black, left=of q_0] {$\top$};
		\node[state] (q_3) [red, draw=black, below=of q_0] {$\bot$};
		
		\path[->] 
		(q_0)
		 edge [bend right, swap, pos=0.6] node {$a$} (q_1)
		 edge [bend left] node {$c$} (q_2)
		 edge [pos=0.7] node {$b,d$} (q_3)
		(q_1) 
		 edge [bend right, swap] node {$b$} (q_0)
		 edge [out=-90, in=0] node[pos=0.8,yshift=0.2cm] {$a,c,d$} (q_3)
		(q_2)
	     edge [bend left] node {$d$} (q_0)
		 edge [out=-90, in=180, swap] node[pos=0.8,yshift=0.2cm] {$a,b,c$} (q_3);
	\end{tikzpicture}
 	\caption{The transducer $\cT_1$ in the proof of~\cref{thm:equivalence_PSPACE-H}.}
 	\label{fig:PSPACE_reduction_T1}
 \end{minipage}%
 \hfill%
 \begin{minipage}{.65\linewidth}
	\begin{tikzpicture}[shorten >=1pt,node distance=1.5cm and 1.2cm,on grid,auto]
		\tikzstyle{state}=[circle, inner sep=0mm, outer sep=0mm, minimum size=8mm, draw=black];
		\node[state] (q_0) {$q$}; 
		\node[state] (q_1) [above right=of q_0] {$q^{0,0}$}; 
		\node[state] (q_2) [right=of q_0] {$q^{0,1}$}; 
		\node[state] (q_3) [above left=of q_0] {$q^{1,0}$}; 
		\node[state] (q_4) [left=of q_0] {$q^{1,1}$}; 
		
		\path[->] 
		(q_0)
		 edge node [pos=0.6] {$0$} (q_1)
		 edge node [pos=0.6] {$0$} (q_2)
		 edge node [pos=0.6, swap] {$1$} (q_3)
		 edge node [pos=0.6, swap] {$1$} (q_4);
	\end{tikzpicture}%
 	\hfill%
	\begin{tikzpicture}[shorten >=1pt,node distance=1.5cm and 1.2cm,on grid,auto]
	    \tikzstyle{state}=[circle, inner sep=0mm, outer sep=0mm, minimum size=8mm, draw=black];
		\node[state] (q_0) {$q$}; 
		\node[state, label={[font=\footnotesize,label distance=-.1cm]above:$q_a$}] (q_1) [above right=of q_0] {$\color{red} \top$}; 
		\node[state, label={[font=\footnotesize,label distance=-.1cm]above:$q_b$}] (q_2) [right=of q_0] {$\color{red} \top$}; 
		\node[state, label={[font=\footnotesize,label distance=-.1cm]above:$q_c$}] (q_3) [above left=of q_0] {$\color{red} \top$}; 
		\node[state, label={[font=\footnotesize,label distance=-.1cm]above:$q_d$}] (q_4) [left=of q_0] {$\color{red} \top$}; 
		\node[state, label={[font=\footnotesize,label distance=-.1cm]above:$q^{0,0}$}] (q_1b)[right=of q_1] {$\color{red} \top$}; 
		\node[state, label={[font=\footnotesize,label distance=-.1cm]above:$q^{0,1}$}] (q_2b)[right=of q_2] {$\color{red} \top$}; 
		\node[state, label={[font=\footnotesize,label distance=-.1cm]above:$q^{1,0}$}] (q_3b)[left=of q_3] {$\color{red} \top$}; 
		\node[state, label={[font=\footnotesize,label distance=-.1cm]above:$q^{1,1}$}] (q_4b)[left=of q_4] {$\color{red} \top$}; 
		
		\path[->] 
		(q_0)
		 edge node [pos=0.6] {$a$} (q_1)
		 edge node [pos=0.6] {$b$} (q_2)
		 edge node [pos=0.6, swap] {$c$} (q_3)
		 edge node [pos=0.6, swap] {$d$} (q_4)
		(q_1) edge node [pos=0.6] {$b$} (q_1b)
		(q_2) edge node [pos=0.6] {$a$} (q_2b)
		(q_3) edge node [pos=0.6, swap] {$d$} (q_3b)
		(q_4) edge node [pos=0.6, swap] {$c$} (q_4b);
	\end{tikzpicture}
 	\caption{Every state and its 4 transitions in $\cN$ (left) turn into 8 transitions in $\cT_2$ (right). All transitions not drawn in the right figure lead to $q_\bot$, a sink state labelled $\color{red}\bot$.}
 	\label{fig:PSPACE_reduction_T2}
 \end{minipage}
\end{figure}

We now proceed to define the reduction formally. We construct $\cT_1$ independently of $\cN$, as depicted in~\cref{fig:PSPACE_reduction_T1}, containing 4 states. For every $x\in \fL$ we have $\cT_1(x)=\top^{|x|}$, and for every other $x\notin \fL$ we have $\cT_1(x)=\top^{m}\bot^{|x|-m}$ where $m$ is the length of the maximal prefix of $x$ in $(ab+cd)^*(a+c+\epsilon)$.

We proceed to construct $\cT_2$. 
We can think of the outgoing transitions from every state $q$ as $\delta(q,0)=\{q^{0,0},q^{0,1}\}$ and $\delta(q,1)=\{q^{1,0},q^{1,1}\}$ (unless $\cN$ has no outgoing transitions on one of the letters, see below). We obtain $\cT_2$ from $\cN$ by introducing 4 new states $q_a,q_b,q_c,q_d$ for every state $q\in Q$, and setting the transitions and labels as depicted in~\cref{fig:PSPACE_reduction_T2}. In case $\cN$ does not have a transition on e.g., $0$ from $q$, then instead of going to $q_a$ or $q_b$, we proceed to a new state $q_\bot$ labelled $\bot$, which is a sink state. In addition, $q_\bot$ is reached upon any transition not yet defined.
Observe that for every $x\in \fL$ we have $\cT_2(x)=\top^{m}\bot^{|x|-m}$ for some $0\le m\le |x|$ (since $q_\bot$ is a sink).

We now claim that $L(\cN)=\{0,1\}^*$ iff $\cT_1\equiv_{2,\fL}\cT_2$.
For the first direction, assume $L(\cN)=\{0,1\}^*$. Observe that $\cT_2\prec_{2,\fL}\cT_1$ independently: for every $x\in (ab+cd)^*$, denote $\cT_2(x)=\top^{m}\bot^{|x|-m}$, then we can construct $x'\req[2]x$ such that $\cT_1(x')=\top^{m}\bot^{|x|-m}$ by leaving $x$ unchanged $m$ steps, and then permuting the letters such that the run of $\cT_1$ moves to the sink labelled $\bot$ (indeed, observe that $m$ must be even by the construction of $\cT_2$, and hence $\cT_1$ can permute e.g., $ab$ to $ba$ in order to start outputting $\bot$ on an even step).

Next, we show that $\cT_1\prec_{2,\fL}\cT_2$. Consider $x\in (ab+cd)^*$, so that $\cT_1(x)=\top^{|x|}$, and let $w\in \{0,1\}^*$ be the word obtained from $x$ by identifying $ab$ with $0$ and $cd$ with $1$. Since $L(\cN)=\{0,1\}^*$, there exists a run (and hence an accepting run) of $\cN$ on $w$, denoted $s_0,s_1,\ldots,s_n$. We now obtain $x''\req[2]x$ by identifying each letter $0$ in $x$ with either $ab$ or $ba$, and each letter $1$ with $cd$ or $dc$, such that the run of $\cT_2$ on $x''$ simulates the run of $\cN$ on $w$. Thus, $\cT_2(x'')=\top^{|x''|}$, and $\cT_2(x'')\req[2]\cT_1(x)$, so we are done.

Conversely, if $\cT_1\equiv_{2,\fL}\cT_2$, then in particular $\cT_1\prec_{2,\fL}\cT_2$. We claim that $L(\cN)=\{0,1\}^*$. Consider $w\in \{0,1\}^*$. Dually to the above, we obtain from $w$ a word $x\in (ab+cd)^*$ by identifying $0$ with $ab$ and $1$ with $cd$, so that $\cT_1(x)=\top^{|x|}$. 
Since $\cT_1\prec_{2,\fL}\cT_2$, there exists $x'\req[2]x$ such that $\cT_2(x')=\top^{|x|}$. Observe that $x'$ must be obtained from $x$ by (possibly) changing each $ab$ to $ba$ and each $cd$ to $dc$. In particular, the run of $\cT_2$ on $x'$ induces a run of $\cN$ on $w$ by identifying both $ab$ and $ba$ as 0 and both $cd$ and $dc$ as 1. This gives $w\in L(\cN)$, so $L(\cN)=\{0,1\}^*$, which concludes the proof. \qed

\subsection{Proof of Theorem~\ref{thm:existential_equivalence_PSPACE-H}}
\label{apx:proof_existential_PSPACE-H}

In order to show that existential round equivalence is $\PSPACE$-hard, we build upon the reduction in the proof of Theorem~\ref{thm:equivalence_PSPACE-H}: we again show a reduction from the universality problem for \NFAs over alphabet $\{0,1\}$ where all states are accepting and the degree of nondeterminism is at most 2 (cf.~\cref{lem:universalityofnfa}).

Consider an \NFA $\cN=\tup{Q,\{0,1\},\delta,q_0,Q}$ where $|\delta(q,\sigma)|\le 2$ for every $q\in Q$ and $\sigma\in \{0,1\}$.
We construct two transducers $\cT_1$ and $\cT_2$ over input and output alphabets $\sI=\{a,b,c,d,\#\}$ and $\sO=\{\top,\bot\}$ and $\fL\subseteq \sI^*$, such that $L(\cN)=\{0,1\}^*$ iff $\cT_1\equiv_{2,\fL}\cT_2$. 

Intuitively, the idea is to use a similar encoding of $\{0,1\}$ in $\{a,b,c,d\}$ whereby $0$ corresponds to either $ab$ or $ba$ and $1$ to $cd$ or $dc$. Now, however, since $k$ is not fixed to $2$, we also allow arbitrary padding with sequences of $\#\#$.

Set $\fL=(ab+cd+\#\#)^*$ (given as a 5 state \DFA). We construct $\cT_1$ and $\cT_2$ similarly to the proof of~\cref{thm:equivalence_PSPACE-H}, by adding self-cycles of length 2 upon reading $\#\#$, from every state except the sink $q_\bot$. See~\cref{fig:existential_PSPACE_reduction_T1,fig:existential_PSPACE_reduction_T2} for an illustration.

\begin{figure}[ht]
	\centering
	\begin{tikzpicture}[shorten >=1pt,node distance=1.5cm and 1.6cm,on grid,auto]
	\tikzstyle{smallnode}=[circle, inner sep=0mm, outer sep=0mm, minimum size=8mm, draw=black];
		\node[smallnode] [initial text={}] (q_0) [initial above] {$\color{red} \top$}; 
		\node[smallnode] (q_1) [right=of q_0] {$\color{red} \top$}; 
		\node[smallnode] (q_2) [left=of q_0] {$\color{red} \top$};
		\node[smallnode] (q_3) [below=of q_0] {$\color{red} \top$};
		\node[smallnode] (q_4) [below=of q_3] {$\color{red} \bot$};
		
		\path[->] 
		(q_0)
		 edge [bend right, swap, pos=0.6] node {$a$} (q_1)
		 edge [bend left, pos=0.6] node {$c$} (q_2)
		 edge [bend left=10] node {$\#$} (q_3)
		 edge [out=-135, in=135, swap, pos=0.7] node {$b,d$} (q_4)
		(q_1) 
		 edge [bend right, swap] node {$b$} (q_0)
		 edge [out=-45, in=-45] node {$a,c,d,\#$} (q_4)
		(q_2)
	     edge [bend left] node {$d$} (q_0)
		 edge [out=-135, in=-135, swap] node {$a,b,c,\#$} (q_4)
		(q_3)
		 edge [bend left=10] node {$\#$} (q_0)
		 edge [] node {$a,b,c,d$} (q_4);
		 
	\end{tikzpicture}
	\caption{The transducer $\cT_1$ in  the proof of~\cref{thm:existential_equivalence_PSPACE-H}.}
	\label{fig:existential_PSPACE_reduction_T1}
\end{figure}

\begin{figure}[ht]
	\centering
	\begin{tikzpicture}[shorten >=1pt,node distance=1.5cm and 1.6cm,on grid,auto]
		\tikzstyle{state}=[circle, inner sep=.5mm, outer sep=0mm, minimum size=8mm, draw=black];
		\node[state] (q_0) {$q$}; 
		\node[state] (q_1) [above right=of q_0] {$q^{0,0}$}; 
		\node[state] (q_2) [right=of q_0] {$q^{0,1}$}; 
		\node[state] (q_3) [above left=of q_0] {$q^{1,0}$}; 
		\node[state] (q_4) [left=of q_0] {$q^{1,1}$}; 
		
		\path[->] 
		(q_0)
		 edge node [pos=0.6] {$0$} (q_1)
		 edge node [pos=0.6] {$0$} (q_2)
		 edge node [pos=0.6,swap] {$1$} (q_3)
		 edge node [pos=0.6,swap] {$1$} (q_4);
	\end{tikzpicture}%
	\hspace{2cm}%
	\begin{tikzpicture}[shorten >=1pt,node distance=1.5cm and 1.6cm,on grid,auto]
	    \tikzstyle{state}=[circle, inner sep=.5mm, outer sep=0mm, minimum size=8mm, draw=black];
		\node[state] (q_0) {$q$}; 
		\node[state] (q') [above=of q_0] {$q_\#$}; 
		\node[state, label={[font=\small]above:$q_a$}] (q_1) [above right=of q_0] {$\color{red} \top$}; 
		\node[state, label={[font=\small]above:$q_b$}] (q_2) [right=of q_0] {$\color{red} \top$}; 
		\node[state, label={[font=\small]above:$q_c$}] (q_3) [above left=of q_0] {$\color{red} \top$}; 
		\node[state, label={[font=\small]above:$q_d$}] (q_4) [left=of q_0] {$\color{red} \top$}; 
		\node[state, label={[font=\small]right:$q^{0,0}$}] (q_1b)[right=of q_1] {$\color{red} \top$}; 
		\node[state, label={[font=\small]right:$q^{0,1}$}] (q_2b)[right=of q_2] {$\color{red} \top$}; 
		\node[state, label={[font=\small]left:$q^{1,0}$}] (q_3b)[left=of q_3] {$\color{red} \top$}; 
		\node[state, label={[font=\small]left:$q^{1,1}$}] (q_4b)[left=of q_4] {$\color{red} \top$}; 
		
		\path[->] 
		(q_0)
		 edge node [pos=0.8] {$a$} (q_1)
		 edge node [pos=0.8] {$b$} (q_2)
		 edge node [pos=0.8, swap] {$c$} (q_3)
		 edge node [pos=0.8, swap] {$d$} (q_4)
		 edge [bend left=10, pos=0.6] node {$\#$} (q')
		(q_1) edge node [pos=0.6] {$b$} (q_1b)
		(q_2) edge node [pos=0.6] {$a$} (q_2b)
		(q_3) edge node [pos=0.6, swap] {$d$} (q_3b)
		(q_4) edge node [pos=0.6, swap] {$c$} (q_4b)
		(q') edge [bend left=10, pos=0.5] node {$\#$} (q_0);
	\end{tikzpicture}
	\caption{Every state and its 4 transitions in $\cN$ (left) turn into 10 transitions in $\cT_2$ (right). All transitions not drawn in the right figure lead to $q_\bot$, a sink state labelled $\color{red}\bot$.}
	\label{fig:existential_PSPACE_reduction_T2}
\end{figure}

We claim that $L(\cN)=\{0,1\}^*$ iff there exists $k>0$ such that $\cT_1\equiv_{k,\fL}\cT_2$.
For the first direction, assume $L(\cN)=\{0,1\}^*$, then we can show that $\cT_1\equiv_{2,\fL}\cT_2$ by following the proof of~\cref{thm:equivalence_PSPACE-H} line for line, with the addition that blocks of the form $\#\#$ leave the state of both $\cT_1$ and $\cT_2$ unchanged.

For the converse direction, assume $\cT_1\equiv_{k,\fL}\cT_2$, and in fact we only assume $\cT_1\prec_{k,\fL}\cT_2$ for some $k>0$. We further assume w.l.o.g.\! that $k$ is even, otherwise we can just take $2k$ (since we also have $\cT_1\prec_{2k,\fL}\cT_2$).

Consider $w\in \{0,1\}^*$. We obtain from $w$ a word $x\in (ab+cd+\#\#)^*$ by identifying $0$ with $ab\#^{k-2}$ and $1$ with $cd\#^{k-2}$. Observe that $\cT_1(x)=\top^{|x|}$, and that $x$ is indeed a $k$-round word in $\fL$, with each round being either $ab\#^{k-2}$ or $cd\#^{k-2}$. 

Since $\cT_1\prec_{k,\fL}\cT_2$, there exists $x'\req[k]x$ such that $\cT_2(x')=\top^{|x|}$. Observe that $x'$ must be obtained from $x$ by (possibly) changing each $ab$ to $ba$ and each $cd$ to $dc$, and by shifting the location of this pair within the $\#$ symbols. Indeed, otherwise the run of $\cT_2$ on $x'$ ends in $q_{\bot}$.
In particular, the run of $\cT_2$ on $x'$ induces a run of $\cN$ on $w$ by identifying both $ab$ and $ba$ as 0 and both $cd$ and $dc$ as 1. Thus, $w\in L(\cN)$, so $L(\cN)=\{0,1\}^*$, and the proof is concluded. \qed

\section{Variants of Round Simulation}
\label{apx:variants}

We start by presenting some transducers that aid us in the proof of strictness of the remaining notions, all being variants of RR:
\begin{enumerate}
    \item RR that expects all requests in the beginning of every round, but outputs like the original (e.g. $\{0,2\}\{1\}\{1\}$ would output $\{0\}\emptyset\{2\}$), modelled by $\cT_1$.
    \item RR that expects input as in the original, but outputs all grants in the end of the round (e.g. $\{0,2\}\{1\}\{1\}$ would output $\emptyset\emptyset\{0,1\}$), modelled by $\cT_2$.
    \item RR such that every other round begins by considering requests of Process 1 before Process 0 (e.g. $\{0\}\{1\}\emptyset\cdot\{0\}\{1\}\emptyset$ would output $\{0\}\emptyset\emptyset\cdot\emptyset\{1\}\emptyset$), modelled by $\cT_3$.
\end{enumerate}
Denote by $\cT$ the transducer for RR. It is not difficult to see that $\cT_1\prectype{s}{u} \cT$ but $\cT_1\not\prectype{\ell}{u} \cT$; that $\cT_2\prectype{u}{s} \cT$ but $\cT_2\not\prectype{u}{\ell} \cT$; and that $\cT_3\prectype{\ell}{\ell} \cT$ but $\cT_3\not\prectype{\ell}{u} \cT$ and $\cT_3\not\prectype{u}{\ell} \cT$.

\begin{exa}
\label{apx:example_counterex}
The transducers in~\cref{fig:example_counterex} satisfy $\cT_1\prectype{\ell}{s}_2 \cT_2$ and $\cT_1\prectype{s}{\ell}_2 \cT_2$. This is proved in~\cref{tab:example_counterex}, which considers all possible forms of each round and gives round equivalent words $x^\rstype{\ell}\reqtype[2]{\ell} x$ and $x^\rstype{s}\reqtype[2]{s} x$ that satisfy the requirements of the definitions.

\begin{figure}[ht]
    \centering
	\begin{tikzpicture}[shorten >=1pt,node distance=1.3cm and 3.5cm,on grid,auto]
	    \tikzset{every state/.style={minimum size=9mm, inner sep=0}};
	    \node[state] (q_0)  [initial, fill=orange!20] {$q_0$};
	    \node[state] (s_0)  [right=of q_0, text=red] {\small $\emptyset$};
	    \node[state] (s_1)  [below=of s_0, text=red] {\small $\emptyset$};
	    \node[state] (s_2)  [below=of s_1, text=red] {\small $\emptyset$};
	    \node[state] (s_3)  [below=of s_2, text=red] {\small $\emptyset$};
	    \node[state] (p_0)  [right=of s_0, text=red, fill=orange!20, label={right:{\footnotesize\color{blue!80} behave like} $q_0$}] {\footnotesize $\{0,1\}$};
	    \node[state] (p_1)  [right=of s_3, text=red, fill=orange!20, label={right:{\footnotesize\color{blue!80} behave like} $q_0$}] {\small $\{0\}$};
        
        \path[->] 
        (q_0) edge [sloped, out=0, in=180] node {$\emptyset$} (s_0)
        (q_0) edge [sloped, out=-30, in=180] node {$\{0,1\}$} (s_1)
        (q_0) edge [sloped, out=-60, in=180] node {$\{0\}$} (s_2)
        (q_0) edge [sloped, out=-90, in=180] node {$\{1\}$} (s_3)
        (s_0) edge [sloped, out=0, in=180, pos=0.8] node {$\{0,1\}$} (p_0)
        (s_1) edge [sloped, out=15, in=-150, pos=0.8] node {$\emptyset$} (p_0)
        (s_2) edge [sloped, out=15, in=-120, pos=0.8] node {$\{1\}$} (p_0)
        (s_3) edge [sloped, out=30, in=-90, pos=0.8] node {$\{0\}$} (p_0)
        (s_0) edge [sloped, out=-30, in=90, pos=0.8] node {else} (p_1)
        (s_1) edge [sloped, out=-15, in=120, pos=0.8] node {else} (p_1)
        (s_2) edge [sloped, out=-15, in=150, pos=0.8] node {else} (p_1)
        (s_3) edge [sloped, out=0, in=180, pos=0.8] node {else} (p_1);
    \end{tikzpicture}
	\begin{tikzpicture}[shorten >=1pt,node distance=1.3cm and 3.5cm,on grid,auto]
	    \tikzset{every state/.style={minimum size=9mm, inner sep=0}};
	    \node[state] (q_0)  [initial, fill=orange!20] {$q_0$};
	    \node[state] (s_0)  [right=of q_0, text=red] {\small $\{0\}$};
	    \node[state] (s_1)  [below=of s_0, text=red] {\small $\{0\}$};
	    \node[state] (s_2)  [below=of s_1, text=red] {\small $\emptyset$};
	    \node[state] (s_3)  [below=of s_2, text=red] {\small $\emptyset$};
	    \node[state] (p_0)  [right=of s_0, text=red, fill=orange!20, label={right:{\footnotesize\color{blue!80} behave like} $q_0$}] {\footnotesize $\{1\}$};
	    \node[state] (p_1)  [right=of s_1, text=red, fill=orange!20, label={right:{\footnotesize\color{blue!80} behave like} $q_0$}] {\small $\emptyset$};
	    \node[state] (p_2)  [right=of s_2, text=red, fill=orange!20, label={right:{\footnotesize\color{blue!80} behave like} $q_0$}] {\footnotesize $\{0,1\}$};
	    \node[state] (p_3)  [right=of s_3, text=red, fill=orange!20, label={right:{\footnotesize\color{blue!80} behave like} $q_0$}] {\small $\{0\}$};
        
        \path[->] 
        (q_0) edge [sloped, out=0, in=180] node {$\emptyset$} (s_0)
        (q_0) edge [sloped, out=-30, in=180] node {$\{0,1\}$} (s_1)
        (q_0) edge [sloped, out=-60, in=180] node {$\{0\}$} (s_2)
        (q_0) edge [sloped, out=-90, in=180] node {$\{1\}$} (s_3)
        (s_0) edge [sloped, pos=0.8, bend left=15] node {$\{0,1\}$} (p_0)
        (s_1) edge [sloped, pos=0.8] node {$\emptyset$} (p_0)
        (s_2) edge [sloped, pos=0.8, bend left=15] node {$\{1\}$} (p_2)
        (s_3) edge [sloped, pos=0.8] node {$\{0\}$} (p_2)
        (s_0) edge [sloped, pos=0.8] node {else} (p_1)
        (s_1) edge [sloped, pos=0.8, bend right=15] node {else} (p_1)
        (s_2) edge [sloped, pos=0.8] node {else} (p_3)
        (s_3) edge [sloped, pos=0.8, bend right=15] node {else} (p_3);
    \end{tikzpicture}
	\caption{Transducers $\cT_1$ (up) and $\cT_2$ (down) in~\cref{apx:example_counterex}, satisfying $\cT_1\prectype{\ell}{s}_2 \cT_2$ and $\cT_1\prectype{s}{\ell}_2 \cT_2$, but $\cT_1\not\prectype{\ell}{\ell}_2 \cT_2$. See~\cref{tab:example_counterex} for a table summarizing the possible inputs and outputs for $\cT_1$.}
    \label{fig:example_counterex}
\end{figure}

However, $\cT_1\not\prectype{\ell}{\ell}_{k'} \cT_2$ for any $k'>0$. Indeed, consider the word $x=\{0,1\}\emptyset^{k'-1}$ having output $\cT_1(x)=\emptyset\{0,1\}\emptyset^{k'-2}$. For $\cT_2$ to output the letter $\{0,1\}$, it must see one of the input letters $\{0\}$ and $\{1\}$, since the only state labelled $\{0,1\}$ has two incoming transitions with $\{0\}$ and $\{1\}$. But any $x'\reqtype[{k'}]{\ell} x$ will not contain the letters $\{0\}$ and $\{1\}$, so $\cT_1(x)\not\reqtype[{k'}]{\ell} \cT_2(x')$. Therefore $\cT_1\not\prectype{\ell}{\ell}_{k'} \cT_2$.

\begin{table}[!htb]
    \centering
    \caption{A table summarizing the outputs of transducer $\cT_1$ in~\cref{apx:example_counterex} on words $x$ of length 2, and round equivalent words $x^\rstype{\ell}$ and $x^\rstype{s}$ that satisfy the requirement of $x'$ in the definition of $\cT_1\prectype{\ell}{s}_2 \cT_2$ and $\cT_1\prectype{s}{\ell}_2 \cT_2$.}
    \vspace{2mm}
    \def\arraystretch{1.3}
    \newcolumntype{x}{>{\columncolor{blue!10}}c}
    \small
    \begin{tabular}{|x|c|x|c|x|c|}
\hline
\rowcolor{blue!30}
$x$ & $\cT_1(x)$ & $x^\rstype{s}$ \ST $\cT_2(x^\rstype{s}) \reqtype[2]{p} \cT_1(x)$ & $\cT_2(x^\rstype{s})$ & $x^\rstype{p}$ \ST $\cT_2(x^\rstype{p}) \reqtype[2]{s} \cT_1(x)$ & $\cT_2(x^\rstype{p})$ \\
\hline \hline
$\emptyset\emptyset$ & $\emptyset\{0\}$ & $\emptyset\emptyset$ & $\{0\}\emptyset$ & $\emptyset\emptyset$ & $\{0\}\emptyset$ \\
\hline
$\emptyset\{0\}$ & $\emptyset\{0\}$ & $\emptyset\{0\}$ & $\{0\}\emptyset$ & $\emptyset\{0\}$ & $\{0\}\emptyset$ \\
\hline
$\emptyset\{1\}$ & $\emptyset\{0\}$ & $\emptyset\{1\}$ & $\{0\}\emptyset$ & $\emptyset\{1\}$ & $\{0\}\emptyset$ \\
\hline
$\emptyset\{0,1\}$ & $\emptyset\{0,1\}$ & $\emptyset\{0,1\}$ & $\{0\}\{1\}$ & $\{0\}\{1\}$ & $\emptyset\{0,1\}$ \\
\hline \hline
$\{0\}\emptyset$ & $\emptyset\{0\}$ & $\{0\}\emptyset$ & $\emptyset\{0\}$ & $\{0\}\emptyset$ & $\emptyset\{0\}$ \\
\hline
$\{0\}\{0\}$ & $\emptyset\{0\}$ & $\{0\}\{0\}$ & $\emptyset\{0\}$ & $\{0\}\{0\}$ & $\emptyset\{0\}$ \\
\hline
$\{0\}\{1\}$ & $\emptyset\{0,1\}$ & $\{0\}\{1\}$ & $\emptyset\{0,1\}$ & $\{0\}\{1\}$ & $\emptyset\{0,1\}$ \\
\hline
$\{0\}\{0,1\}$ & $\emptyset\{0\}$ & $\{0\}\{0,1\}$ & $\emptyset\{0\}$ & $\{0\}\{0,1\}$ & $\emptyset\{0\}$ \\
\hline \hline
$\{1\}\emptyset$ & $\emptyset\{0\}$ & $\{1\}\emptyset$ & $\emptyset\{0\}$ & $\{1\}\emptyset$ & $\emptyset\{0\}$ \\
\hline
$\{1\}\{0\}$ & $\emptyset\{0,1\}$ & $\{1\}\{0\}$ & $\emptyset\{0,1\}$ & $\{1\}\{0\}$ & $\emptyset\{0,1\}$ \\
\hline
$\{1\}\{1\}$ & $\emptyset\{0\}$ & $\{1\}\{1\}$ & $\emptyset\{0\}$ & $\{1\}\{1\}$ & $\emptyset\{0\}$ \\
\hline
$\{1\}\{0,1\}$ & $\emptyset\{0\}$ & $\{1\}\{0,1\}$ & $\emptyset\{0\}$ & $\{1\}\{0,1\}$ & $\emptyset\{0\}$ \\
\hline \hline
$\{0,1\}\emptyset$ & $\emptyset\{0,1\}$ & $\{0,1\}\emptyset$ & $\{0\}\{1\}$ & $\{0\}\{1\}$ & $\emptyset\{0,1\}$ \\
\hline
$\{0,1\}\{0\}$ & $\emptyset\{0\}$ & $\{0,1\}\{0\}$ & $\{0\}\emptyset$ & $\{0,1\}\{0\}$ & $\{0\}\emptyset$ \\
\hline
$\{0,1\}\{1\}$ & $\emptyset\{0\}$ & $\{0,1\}\{1\}$ & $\{0\}\emptyset$ & $\{0,1\}\{1\}$ & $\{0\}\emptyset$ \\
\hline
$\{0,1\}\{0,1\}$ & $\emptyset\{0\}$ & $\{0,1\}\{0,1\}$ & $\{0\}\emptyset$ & $\{0,1\}\{0,1\}$ & $\{0\}\emptyset$ \\

\hline
    \end{tabular}
    \label{tab:example_counterex}
\end{table}
\end{exa}

\end{document}